\documentclass[journal]{IEEEtran}

\usepackage{color}
\usepackage{lineno}
\usepackage{cite}
\usepackage{verbatim}
\usepackage{pifont}
\usepackage{longtable}
\usepackage{tabularx}  
\usepackage{ltxtable}  
\usepackage{stfloats}
\usepackage{amsmath,amssymb,amsfonts}
\usepackage{algorithmic}
\usepackage{graphicx}
\usepackage{textcomp}
\usepackage{xcolor}
\usepackage{subfigure}
\usepackage{setspace}
\usepackage{svg}
\usepackage[ruled]{algorithm2e}
\usepackage{booktabs}
\usepackage{url}
\usepackage{diagbox}
\usepackage{xcolor}
\usepackage{multicol}
\usepackage{multirow}
\usepackage{tikz}
\usepackage[colorlinks=true,linkcolor=blue,citecolor=blue,urlcolor=blue]{hyperref}
\definecolor{lime}{HTML}{A6CE39}
\DeclareRobustCommand{\orcidicon}{%
    \begin{tikzpicture}
    \draw[lime, fill=lime] (0,0) 
    circle [radius=0.16] 
    node[white] {{\fontfamily{qag}\selectfont \tiny ID}};    \draw[white, fill=white] (-0.0625,0.095) 
    circle [radius=0.007];    \end{tikzpicture}
    \hspace{-2mm}}
\foreach \x in {A, ..., Z}{%
    \expandafter\xdef\csname orcid\x\endcsname{\noexpand\href{https://orcid.org/\csname orcidauthor\x\endcsname}{\noexpand\orcidicon}}
    }

\SetCommentSty{mycommfont}

\SetKwInput{KwInput}{Input}                
\SetKwInput{KwOutput}{Output}              

\begin{document}



\title{RadHARSimulator V2: Video to Doppler Generator\\
\thanks{Manuscript received XXXXXXX XX, 2025; revised XXXXXXX XX, 2025; accepted XXXXXXX XX, 2025. Date of publication XXXXXXX XX, 2025; date of current version XXXXXXX XX, 2025.\par
My Bio: My name is Weicheng Gao. I'm a Ph.D. student from Beijing Institute of Technology. I’m majored and interested in mathematical and modeling theory research of signal processing, radar signal processing techniques, and AI for radar, apprenticed under professor Xiaopeng Yang. I’m currently dedicated in the field of Through-the-Wall Radar Human Activity Recognition. Looking forward to learning and collaborating with more like-minded teachers and mates. (e-mail: JoeyBG@126.com).\par
Digital Object Identifier 10.48550/arXiv.2511.XXXXX.\par}}

\author{Weicheng~Gao\orcidA{},~\IEEEmembership{Graduate~Student~Member,~IEEE}   
        \vspace{-0.2cm}
        }
        
\markboth{arXiv Preprint, November, 2025}%
{Shell \MakeLowercase{\textit{et al.}}: Bare Demo of IEEEtran.cls for IEEE Journals}

\maketitle

\begin{abstract}
Radar-based human activity recognition (HAR) lies in providing decision-making insights for urban warfare and counter-terrorism through contactless perception, while safeguarding privacy and security in areas such as monitoring for elderly individuals living alone and smart home applications. However, this field still lacks a comprehensive simulation method for radar-based HAR. Existing software is developed based on models or motion-captured data, resulting in limited flexibility. To address this issue, a simulator that directly generates Doppler spectra from recorded video footage (RadHARSimulator V2) is presented in this paper. Both computer vision and radar modules are included in the simulator. In computer vision module, the real-time model for object detection with global nearest neighbor is first used to detect and track human targets in the video. Then, the high-resolution network is used to estimate two-dimensional poses of the detected human targets. Next, the three-dimensional poses of the detected human targets are obtained by nearest matching method. Finally, smooth temporal three-dimensional pose estimation is achieved through Kalman filtering. In radar module, pose interpolation and smoothing are first achieved through the Savitzky-Golay method. Second, the delay model and the mirror method are used to simulate echoes in both free-space and through-the-wall scenarios. Then, range-time map is generated using pulse compression, moving target indication, and DnCNN. Next, Doppler-time map (DTM) is generated using short-time Fourier transform and DnCNN again. Finally, the ridge features on the DTM are extracted using the maximum local energy method. In addition, a hybrid parallel-serial neural network architecture is proposed for radar-based HAR. Numerical experiments are conducted and analyzed to demonstrate the effectiveness of the designed simulator and the proposed network model. The open-source code of this work can be found in: \href{https://github.com/JoeyBGOfficial/RadHARSimulatorV2-Video-to-Doppler-Generator}{GitHub/JoeyBGOfficial/RadHARSimulatorV2}\par
\end{abstract}


\IEEEpeerreviewmaketitle

\section{Introduction}
\IEEEPARstart{I}{n} recent years, as a non-contact sensing technique, radar-based human activity recognition (HAR) has become a research hotspot in both academia and industry \cite{Main1, Main2, Main3, Main4, Main5}. Its main benefit is that it makes it possible for intelligent perception to function continuously without being impacted by lighting or violating people's privacy \cite{CuiGL, JinT, DingYP, YeSB}. This field is rapidly developing from a cutting-edge academic issue to large-scale practical applications, positioned to become an essential weapon in effective warfare and terrorist operations, as well as to build an intelligent, user-friendly, and privacy-secure social infrastructure \cite{WangJQ, HongH, YangY}.\par
\begin{figure}
    \centering
    \includegraphics[width=0.48\textwidth]{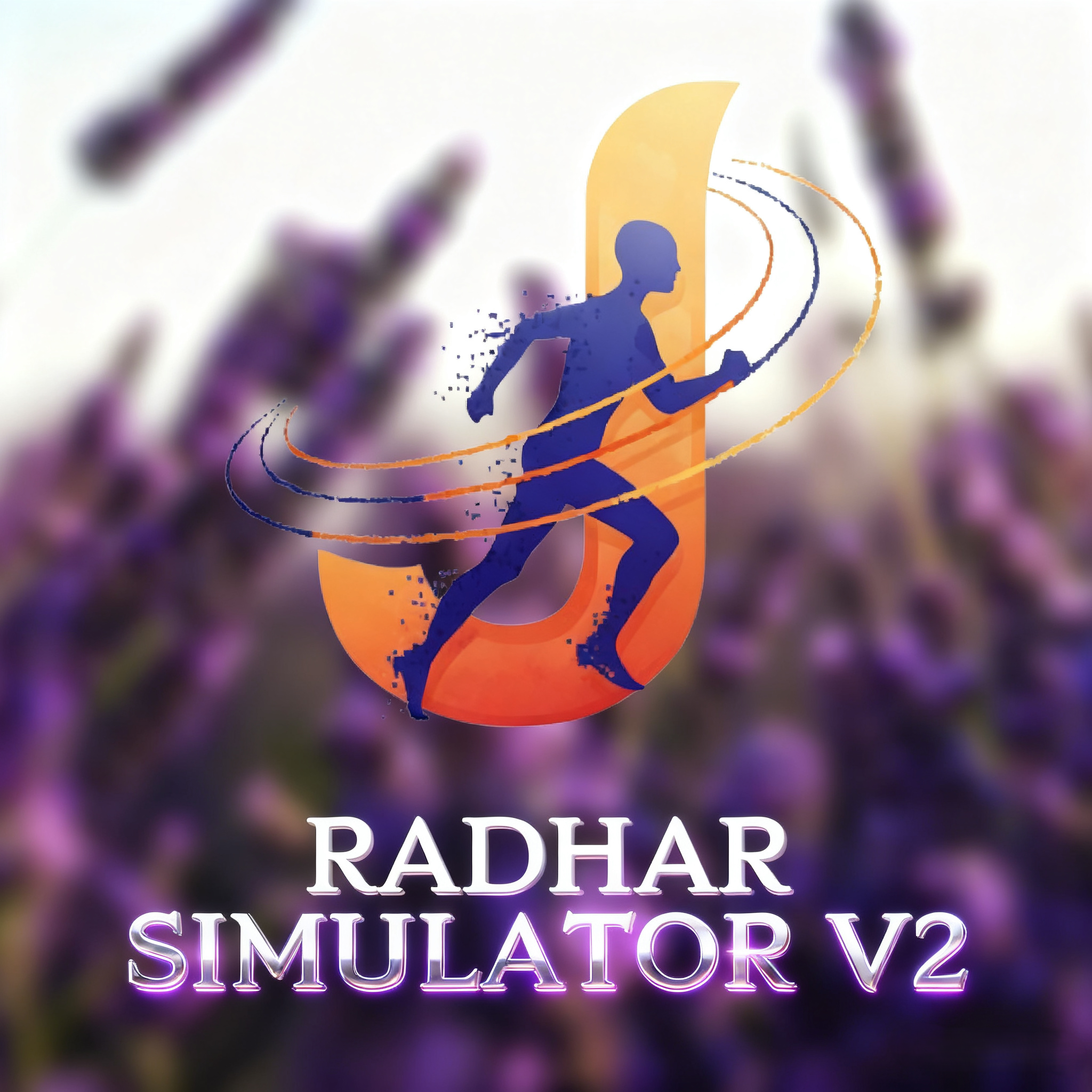}
    \caption{Splash screen of RadHARSimulator V2.}
    \label{Splash_Screen}
    \vspace{-0.4cm}
\end{figure}\par
\begin{figure*}[!ht]
    \centering
    \includegraphics[width=\textwidth]{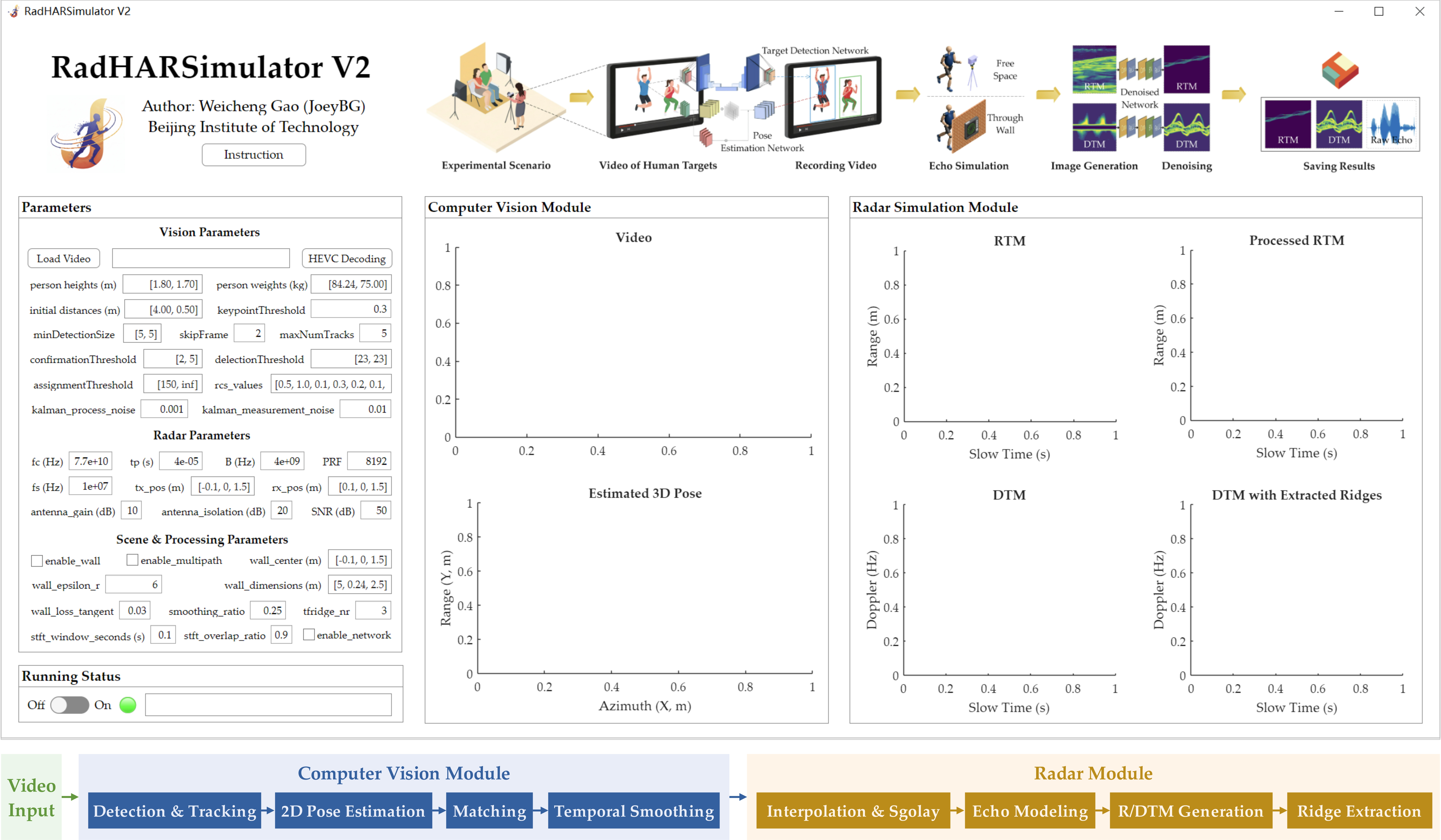}
    \caption{The interface and processing overflow of RadHARSimulator V2.}
    \label{Over_Flow}
    \vspace{-0.3cm}
\end{figure*}\par
Research in radar-based HAR spans over two decades, encompassing progressive levels of development such as human body models, statistical learning methods based on manual feature selection, and deep learning approaches utilizing automatic feature extraction \cite{Survey1}. However, research in this field was accelerated significantly five years ago, with an increasing number of researchers and institutions dedicating resources to developing more precise and cost-effective methods in hopes of enhancing their practical value \cite{Survey2, Survey3}. Among them, a method combining an improved principal component analysis and an improved VGG-16 model was proposed by Zhao et Al. to solve the problem of over-fitting caused by the small-scale dataset, and an accuracy of $96.34\%$ on HAR was achieved \cite{Zhao_ApplSci}. A time-distributed convolutional neural network (CNN) enhanced with a bidirectional long-short term memory (Bi-LSTM) mechanism was used by Singh et Al. to classify five human full-body activities, and an accuracy of $90.47\%$ was achieved \cite{Singh_ACMWirelessNetw}. A real-time human activity recognition method for through-the-wall (TTW) radar was proposed by Cheng et Al. to solve the problem of unknown temporal allocation of activities during recognition, and an average accuracy of $97.6\%$ was achieved \cite{Cheng_IEEERadarConf}. A hybrid CNN–LSTM network was used by Zhu et Al. to classify human activities based on micro-Doppler radar, and an accuracy of $98.28\%$ was achieved \cite{Zhu_IEEEAccess}. The problem that deploying an HAR model in unseen environments remains challenging was solved by El Hail et Al. by applying data augmentation and unsupervised domain adaptation to enhance the robustness of HAR models, and the F1-score was improved by $0.281$ with data augmentation and by $0.337$ when the DANN strategy was jointly used with data augmentation \cite{ElHail_Electronics}. A method of a two-stream one-dimensional CNN with a bidirectional gated recurrent unit was proposed by Tan et Al. to recognize six daily activities, and an impressive accuracy rate of $98.2\%$ was achieved \cite{Tan_Sensors}. A stacked recurrent neural network (RNN) was used by Wang et Al. to recognize human motion, and an accuracy of $92.65\%$ was achieved \cite{Wang_DigitSignalProcess}. A Bi-LSTM network was used by Li et Al. to multimodal continuous HAR and fall detection, and an accuracy of over $95\%$ was achieved \cite{Li_IEEESensJ}. LSTM was used by Noori et Al. for the static actions with a non-wearable ultra-wideband (UWB) radar, and an accuracy approaching $99.6\%$ was achieved \cite{Noori_IEEEAccess}. Frequency-modulated continuous wave radar systems were studied by Pesin et Al. for radar-based HAR, and an average classification accuracy of $89.8\%$ for the millimeter wave radar system and $95.7\%$ for the sub-6 GHz radar was achieved \cite{Pesin_IEEEIntConfConsumElectron}. A deep convolutional autoencoder was used by Seyfioglu et Al. for radar-based classification of similar aided and unaided human activities, and the highest accuracy of $94.2\%$ was achieved \cite{Seyfioglu_IEEETTransAerospElectronSyst}. A multi-frequency spectrograms was used by Ding et Al. to continuous human activity recognition through parallelism LSTM, and accuracies of $85.41\%$ and $96.15\%$ were achieved \cite{Ding_RemoteSens}. An open source pretrained CNN was used by Chakraborty et Al. to train with their own provided dataset, and an overall accuracy of $98\%$ was achieved \cite{Chakraborty_PatternRecognitLett}. A Transformer was trained as an end-to-end model and used for the classification of seven coarse-scaled tasks by Yan et Al., and the accuracy of the Transformer was the highest at $90.45\%$ \cite{Yan_IEEEGlobCommunConf}. A comparative study on recent progress of machine learning-based HAR with radar was performed by Papadopoulos and Jelali, and the CNN-based classification achieved better performance in comparison to the investigated RNN-based methods \cite{Papadopoulos_ApplSci}. Our team also conducted relevant researches, primarily focusing on the TTW scenario, including achieving data augmentation \cite{Gao1}, higher recognition accuracy and robustness \cite{Gao2}, faster inference speeds \cite{Gao3}, improved kinematic models \cite{Gao4}, enhanced feature extraction \cite{Gao5}, and better generalization for targets with varying body types \cite{Gao6}. These works, being focused on the design of intelligent algorithms, invariably required the optimization or training of a certain amount of prior data. For the pre-verification phase of algorithms, generating large amounts of echoes or spectra using simulation methods is a necessary area of research.\par
Not long ago, RadHARSimulator V1, a model-based radar HAR simulator integrating an anthropometrically scaled 13-scatterer kinematic model to simulate 12 activities is developed \cite{RadHARSimulatorV1}, generating radar echoes with dynamic radar cross-section, propagation effects, and noise, processing data through a pipeline to produce range-time map (RTM) and Doppler-time map (DTM). However, shortcomings are exhibited, including limited waveform and radar support, simplistic models omitting complex motions and multipath, and insufficient user testing. Therefore, methods for generating radar echoes and spectra from recorded videos are deemed necessary, as they capture real-world visual cues with pose variations, enhancing simulation realism, diversity, and applicability for challenging HAR beyond analytical models.\par
\begin{figure*}[!ht]
    \centering
    \includegraphics[width=\textwidth]{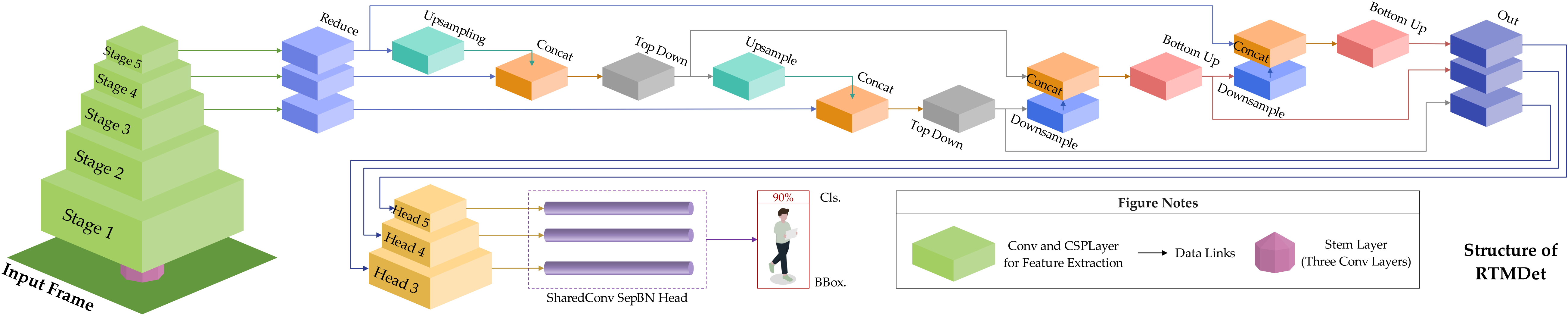}
    \caption{Structure design of the RTMDet for detecting human targets in video frames.}
    \label{RTMDet}
    \vspace{-0.2cm}
\end{figure*}\par
Taking all the above considerations into account, RadHARSimulator V2, a simulator for generating radar echoes and spectra from recorded video footage is proposed in this paper, shown in Fig. \ref{Over_Flow}. Specific contributions of this paper are:\par
\textbf{(1) Three-Dimensional (3D) Human Pose and Location Estimation:} First, real-time model for object detection (RTMDet) is employed for video human detection, while the global nearest neighbor (GNN) method is utilized for human tracking. Subsequently, high-resolution network (HRNet) is applied to perform two-dimensional (2D) pose estimation of detected human targets. Next, the nearest matching method is used to achieve 3D pose estimation of human targets. Finally, the geometric parameters derived from the 3D poses are leveraged to estimate the spatial locations of all human joints.\par
\textbf{(2) Temporal Pose Smoothing:} The Kalman filter method is employed to achieve temporal consistency and smoothing in human 3D pose and location estimation.\par
\textbf{(3) Visual-to-Radar Pose Interaction:} Linear interpolation method and Savitzky-Golay filtering technique are employed to achieve pose interaction between video frame rates and radar pulse repetition frequency (PRF), along with second-order smoothing.\par
\textbf{(4) Radar Spectra Generation and Enhancement:} Both free-space and TTW scenarios are introduced, with radar echo generation and multipath effect simulation implemented based on delay models and mirroring methods. Pulse compression and moving target indication (MTI) methods are employed to generate the RTM, while the short-time Fourier transform (STFT) method is used to generate the DTM. Additionally, the DnCNN model is utilized to perform denoising and micro-Doppler signature enhancement on both the RTM and DTM, and the ridge features on DTM are extracted using the maximum local energy (MLE) method.\par
\textbf{(5) Novel Radar-based HAR Method:} a novel hybrid parallel-serial neural network architecture (SPNet) is proposed for radar-based HAR.\par
Besides, numerical experiments are also conducted in this paper, aiming to evaluate the validity and accuracy of the data generated by the simulator, while simultaneously assessing the effectiveness of the proposed network approach.\par
The rest of the paper is organized as follows. The computer vision module, which achieves video to 3D pose and location estimation, is presented in section II. The radar module, which achieves 3D pose to spectra generation and feature extraction, is presented in section III. The design of hybrid parallel-serial neural network model for radar-based HAR is presented in section IV. Numerical experiments, analysis, and discussions are presented in section V. Finally, the conclusion is given in section VI.\par

\section{Computer Vision Module}
In this section, the RTMDet and GNN-based video human detection and tracking method is first presented. Then, the 2D pose estimation of the detected human targets achieved by HRNet is then discussed. Next, the 3D pose estimation is achieved by nearest matching. Finally, the Kalman filtering method is used to achieve temporal smoothing in human 3D pose and location estimation.\par

\subsection{Detection and Tracking Human Targets}
The detection of human targets in video is achieved by RTMDet. Details can be found in \cite{RTMDet}. As shown in Fig. \ref{RTMDet}, at time $t_f$, the image frame of the input video is represented as a matrix $\mathbf{I}_{t_f} \in \mathbb{R}^{H \times W \times C}$, where $H,W$ are the width and height of the frame, and $C$ is th number of channels. Some of the data augmentation techniques are taken to produce the network input tensor $\mathbf{X}_{\mathrm{in},t_f}\in \mathbb{R}^{H' \times W' \times C}$. The overall architecture of RTMDet consists of three core components, including the backbone, the neck, and the detection head.\par
CSPNeXt is adopted as the foundational building module for the backbone network. The network is initiated with a stem layer, followed by four processing stages. The core of the CSPNeXt module is formed by depthwise separable convolutions that utilize large convolution kernels; this design is aimed at extracting image features efficiently.\par
The neck network is structured on a path aggregation feature pyramid network. Feature maps output from different stages of the backbone network are received and fused through a top-down path followed by a bottom-up path. This process integrates contextual information across multiple scales, thereby enhancing the model’s detection capability for targets of varying sizes. The neck network likewise employs CSPNeXt modules to process the feature maps.\par
\begin{figure}
    \centering
    \includegraphics[width=0.48\textwidth]{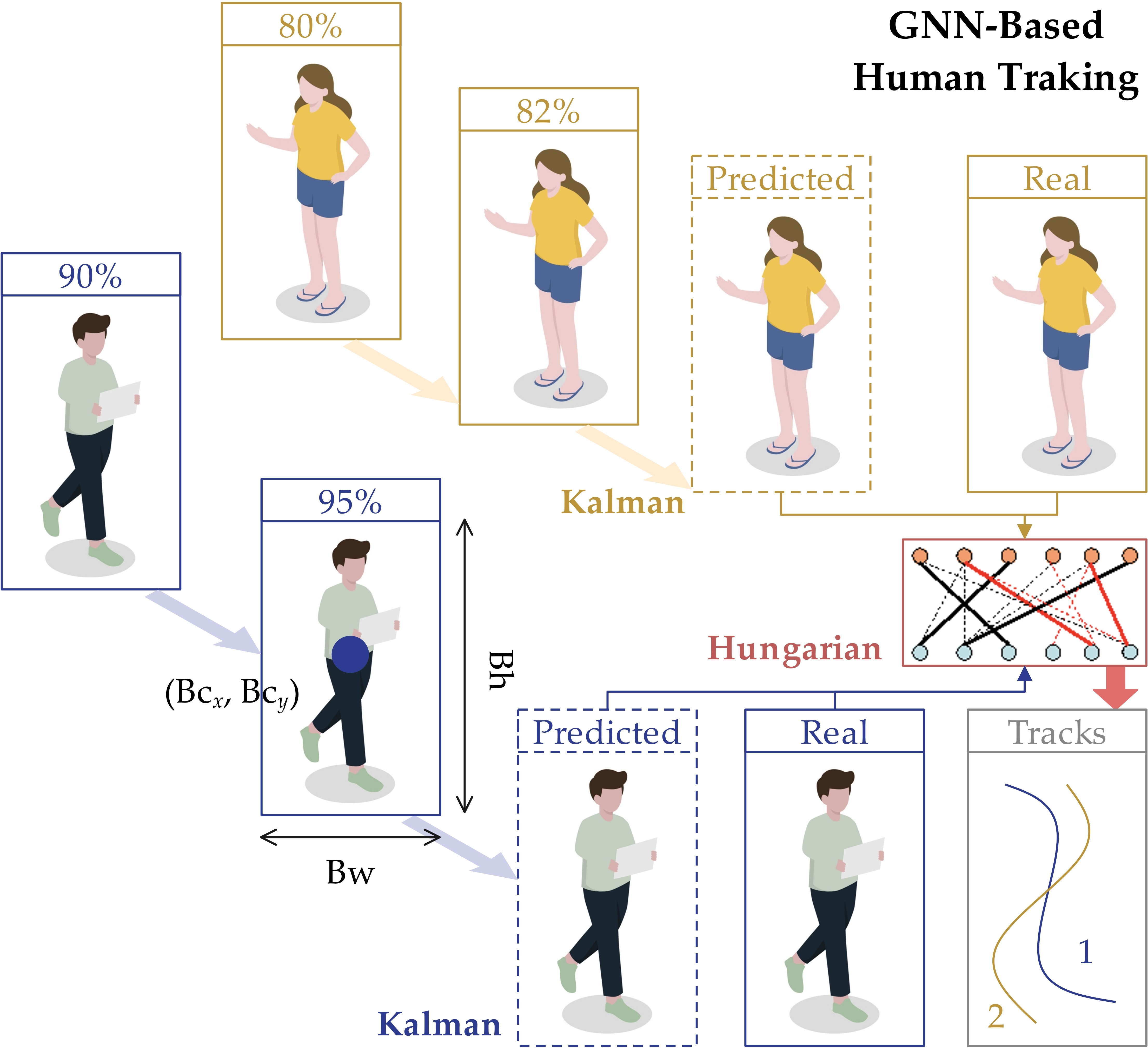}
    \caption{Human tracking based on GNN.}
    \label{GNN}
    \vspace{-0.4cm}
\end{figure}\par
A shared detection head is employed by RTMDet to simultaneously handle classification and regression tasks. To enhance efficiency and reduce parameter count, convolutional layer weights are shared across detection heads at different feature levels, while independent batch normalization layers are maintained for each. This design enables the model to effectively predict object categories and bounding box positions from the fused multi-scale feature maps.\par
The loss function of RTMDet is composed of two components, including classification loss and bounding box regression loss. Positive and negative samples are determined through a dynamic soft label assignment strategy, and the corresponding losses are computed accordingly. The classification task employs quality focal loss (QFL) $\mathcal{L}_\mathrm{QFL}(\cdot)$. Bounding-box localization is performed using generalized intersection over union loss (GIoU) $\mathcal{L}_\mathrm{GIoU}(\cdot)$. Complete loss function $\mathcal{L}_\mathrm{RTMDet}$ is denoted as follows:
\begin{equation}
\begin{aligned}
\mathcal{L}_\mathrm{RTMDet} &= \frac{1}{N_\mathrm{pos}}\sum_{i}\mathcal{L}_\mathrm{QFL}(\mathrm{Sigm}_i,\mathrm{IoU}_i) \\ &+\frac{\lambda_\mathrm{RTMDet}}{N_\mathrm{pos}}\sum_{i}\mathbb{I}_{y_i>0}\cdot \mathcal{L}_\mathrm{GIOU}(\mathrm{BBox}_i,\mathrm{GTB}_i)
\end{aligned},
\end{equation}
where $\sum_i$ is the summation over all prediction samples, $N_\mathrm{pos}$ represents the total number of samples identified as positive by the passive allocation strategy, $\lambda_\mathrm{RTMDet}$ is the weight coefficient for the regression loss, $\mathbb{I}$ is the indicator function, $\mathrm{Sigm}_i$ is the output of the $i^{\mathrm{th}}$ prediction sample after activation by the Sigmoid function, $\mathrm{IoU}_i$ is the true label IoU for the $i^{\mathrm{th}}$ prediction sample, $\mathrm{BBox}_i$ is the bounding box coordinates representing the prediction for the $i^{\mathrm{th}}$ positive sample, and $\mathrm{GTB}_i$ is the coordinates of the true target bounding box matched with the $i^{\mathrm{th}}$ positive sample.\par
At frame $t_f$, the RTMDet outputs a set of bounding box matrices $\mathbf{X}_{\mathrm{Box},t_f} = [{\mathbf{X}_{\mathrm{Box},1,t_f}, \mathbf{X}_{\mathrm{Box},2,t_f}, \dots, \mathbf{X}_{\mathrm{Box},P,t_f}}]$, where each $\mathbf{X}_{\mathrm{Box},p,t_f},~p\in \mathbb{Z}^{+} \cap [1,P]$ corresponds to a detected potential $p^{\mathrm{th}}$ human target, which will be used for subsequent GNN tracking.\par
Assume at $t_f-\Delta t_f$, $M$ valid tracks have been recorded: $\mathbf{Tr}_{t_f-\Delta t_f}=[\mathbf{Tr}_{1,t_f-\Delta t_f},\mathbf{Tr}_{2,t_f-\Delta t_f},\ldots,\mathbf{Tr}_{M,t_f-\Delta t_f}]$. The state of each track $\mathbf{Tr}_{m,t_f-\Delta t_f}, ~m\in \mathbb{Z}^{+} \cap [1,M]$ can be represented by a vector $\mathbf{st}_m,~m\in \mathbb{Z}^{+} \cap [1,M]$ including both position and velocity information, defined as:
\begin{equation}
\mathbf{st}_m=[\mathrm{Bc}_x,\mathrm{Bc}_y,\mathrm{Bw},\mathrm{Bh}, \dot{\mathrm{Bc}}_x,\dot{\mathrm{Bc}}_y,\dot{\mathrm{Bw}},\dot{\mathrm{Bh}}]^{\top},
\end{equation}
where $\mathrm{Bc}_x,\mathrm{Bc}_y$ are center coordinates of the bounding box, $\mathrm{Bw},\mathrm{Bh}$ are the width and height of the bounding box, respectively, and $\dot{}$ means the first derivative of a variable. To match these tracks with the current frame's detection, each track's position at the current frame $t_f$ must first be predicted. Here, a linear motion model is adopted:
\begin{equation}
\hat{\mathbf{st}}_{m,t_f}=\mathbf{TransF} \cdot \mathbf{st}_{m,t_f-\Delta t_f},
\end{equation}
where $\mathbf{TransF}$ is the state transition matrix:
\begin{equation}
\mathbf{TransF} = 
\begin{bmatrix}
\mathbf{I}_{4 \times 4} & \Delta t_f \cdot \mathbf{I}_{4 \times 4} \\
\mathbf{0}_{4 \times 4} & \mathbf{I}_{4 \times 4}
\end{bmatrix},
\end{equation}
where $\mathbf{I}$ is the identity matrix. Through this, the prediction bounding box $\hat{\mathbf{X}}_{\mathrm{Box},m,t_f},~m\in \mathbb{Z}^{+} \cap [1,M]$ for each track in the current frame is obtained.\par
Calculate the IoU between the predicted bounding box $\hat{\mathbf{X}}_{\mathrm{Box},m,t_f}$ and the detected bounding box $\mathbf{X}_{\mathrm{Box},p,t_f}$:
\begin{equation}
\begin{aligned}
\mathrm{IoU}(m,p) &= \frac{\mathrm{Area}(\hat{\mathbf{X}}_{\mathrm{Box},m,t_f} \cap \mathbf{X}_{\mathrm{Box},p,t_f})}{\mathrm{Area}(\hat{\mathbf{X}}_{\mathrm{Box},m,t_f} \cup \mathbf{X}_{\mathrm{Box},p,t_f})}\\ \mathrm{cost}_{mp}&=1-\mathrm{IoU}(m,p)
\end{aligned},
\end{equation}
where $\mathrm{Area}(\cdot)$ is the area calculation function, and $\mathrm{cost}_{mp}$ is the cost element. Combine all cost elements to obtain the cost matrix $\mathbf{cost}_{M\times P}$.\par
After constructing the cost matrix, the data association problem is transformed into a classic assignment problem, which can be solved in polynomial time using the Hungarian algorithm to find the optimal solution. The objective of the algorithm is to find a binary allocation matrix $\mathbf{Alloc}_{M\times P}$ where an element $\mathrm{alloc}_{mp}=1$ indicates that track $m$ is matched with detection $p$, and $0$ otherwise. Solve the optimization problem:
\begin{equation}
\begin{gathered}
\min~\sum_{m=1}^{M}\sum_{p=1}^{P}\mathrm{cost}_{mp}\cdot\mathrm{alloc}_{mp}\\
\mathrm{s.t.}~\sum_{m=1}^{M}\mathrm{alloc}_{mp}\leq 1, \forall m\in \mathbb{Z}^{+}\cap[1,M]\\
\quad\sum_{p=1}^{P}\mathrm{alloc}_{mp}\leq 1, \forall p\in \mathbb{Z}^{+}\cap[1,P]
\end{gathered}.
\end{equation}\par
A set of paired matching relationships is obtained. The GNN updates, creates, and deletes tracks based on the assignment results \cite{GNN}: For successfully matched track-detection pairs $(m,p)$, use the detection box $\mathbf{X}_{\mathrm{Box},p,t_f}$ information to update the state vector $\mathbf{st}_m$ of track $\mathbf{Tr}_m$. Simultaneously, reset the track's counter to $0$. Detection boxes not assigned to any existing track may represent newly emerging targets. A tentative track is only confirmed as a valid new track if it can be matched consecutively across subsequent frames. For tracks that fail to match any detection box in the current frame, their counter increments by $1$. When this counter exceeds a preset threshold, the target is deemed to have left the scene or become completely occluded, and the track is deleted.\par
Through iterative cycles of above steps, the GNN algorithm provides stable and continuous tracking for human targets within video sequences. The detection box $\mathbf{X}_{\mathrm{Box}, t_f}$ for each tracked target will be used for subsequent 2D pose estimation.\par
\begin{figure*}
    \centering
    \includegraphics[width=\textwidth]{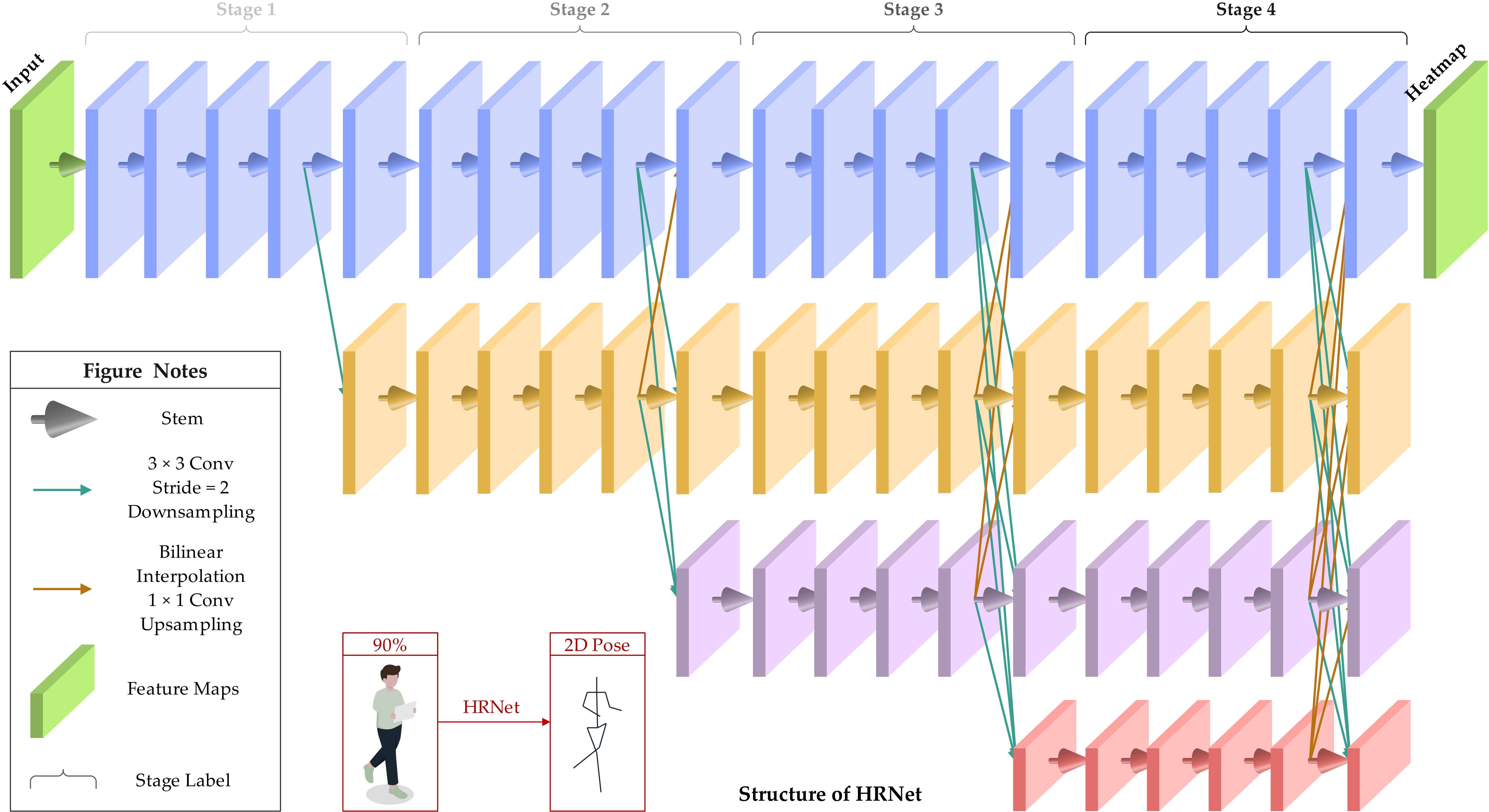}
    \caption{Structure design of the HRNet for estimating 2D poses of the detected human targets.}
    \label{HRNet}
    \vspace{-0.3cm}   
\end{figure*}\par

\subsection{2D Pose Estimation}
The 2D pose estimation of human targets is achieved by HRNet. Details can be found in \cite{HRNet}. As shown in Fig. \ref{HRNet}, high-resolution feature representations are maintained by HRNet throughout the entire network processing, thereby avoiding spatial information loss caused by multiple downsamplings in traditional methods. The input image $\mathbf{X}_{\mathrm{Box}, t_f}$ is used as the starting point, where frame time information can serve as a reference for temporal processing, but the network itself primarily focuses on spatial feature extraction. This objective is achieved by HRNet through the parallel connection of multiple resolution subnets, with the initial stage employing a high-resolution subnet to process the input, and lower-resolution subnets being gradually introduced to capture multi-scale contextual information.\par
The network structure is composed of four stages, each containing a series of convolutional layers and residual blocks. The first stage begins with the high-resolution subnet, where initial features are extracted using convolutional operations. In subsequent stages, lower-resolution subnets are progressively added; these subnets reduce spatial dimensions through strided convolutions while increasing the number of channels to enrich feature representations. A key multi-scale fusion mechanism is performed periodically between different subnets, with information exchanged via upsampling and downsampling operations to ensure effective integration of high-resolution features with low-resolution semantic contexts. This design ensures that detailed information is always preserved in the high-resolution subnet, while global perception is provided by the low-resolution subnets.\par
In the output part, feature maps from all subnets are fused by HRNet, ultimately generating a set of heatmaps, each corresponding to the spatial probability distribution of human joints. These heatmaps are predicted from the fused features through convolutional layers and are directly used to infer the positions of human pose joints.\par
The mean squared error is adopted by the loss function of HRNet, with the objective being to minimize the difference between the joint heatmaps predicted by the network and the ground-truth heatmaps, which is defined as follows:
\begin{equation}
\mathcal{L}_\mathrm{HRNet} = \frac{ \sum_{k=1}^{K} \frac{ \sum_{i=1}^{\mathrm{Bh}} \sum_{j=1}^{\mathrm{Bw}} \left\| \mathbf{Gauss}_{k}(i, j) - \hat{\mathbf{Gauss}}_{k}(i, j) \right\|_2^2}{N_{\mathrm{HRNet},k}}}{K_\mathrm{HRNet}},
\end{equation}
where $K_\mathrm{HRNet}$ is the total number of joints, the number of output joints for HRNet is $17$ in this work, and the number of nodes subsequently used for 3D matching is $14$, $\mathbf{Gauss}_{k}$ is the ground-truth Gaussian heatmap corresponding to the $k^{\mathrm{th}}$ joint, $\hat{\mathbf{Gauss}}_{k}$ is the predicted Gaussian heatmap corresponding to the $k^{\mathrm{th}}$ joint, and $N_{\mathrm{HRNet},k}$ is the normalized factor set to the sum of all pixel values in the ground-truth heatmap.\par
The 2D pose obtained from HRNet inference can be expressed as $\mathbf{P}_\mathrm{2D}\in \mathbb{R}^{K_\mathrm{HRNet}\times 2}$. This result will be used for subsequent mapping from 2D to 3D pose.\par

\subsection{2D to 3D Pose Mathcing}
The nearest matching algorithm is employed to implement the mapping from 2D pose to 3D pose \cite{NearestMatching}. First, extract $1877420$ 3D poses from the Human3.6M training dataset \cite{Human3.6M} and construct the library $\mathbf{Pool}_\mathrm{3D}\in \mathbb{R}^{N_\mathrm{H3.6M}\times K_\mathrm{HRNet}\times 3}$, where $N_\mathrm{H3.6M}$ is the total number of data in the library. The depth axis of the 3D poses in this library is then pooled to obtain their 2D mappings, forming $\mathbf{Pool}_\mathrm{2D}\in \mathbb{R}^{N_\mathrm{H3.6M}\times K_\mathrm{HRNet}\times 2}$. The pose data in the library is derived from diverse human motion and normalized to unit scale.\par
To find the best-matching 3D pose of the $\mathbf{P}_\mathrm{2D}$, the Euclidean distance between $\mathbf{P}_\mathrm{2D}$ and each data in library $\mathbf{Pool}_\mathrm{2D}$ is calculated:
\begin{equation}
d_{\mathrm{pose},i}=|| \mathbf{P}_\mathrm{2D}-\mathbf{Pool}_\mathrm{2D}^{i} ||_2,\quad i\in \mathbb{Z}^{+}\cap [1,N_\mathrm{H3.6M}].
\end{equation}\par
The index $i^*$ of the nearest matching is selected as:
\begin{equation}
i^* = \arg \min_i d_{\mathrm{pose},i}.
\end{equation}\par
The corresponding 3D pose is then retrieved as $\mathbf{P}_\mathrm{3D}=\mathbf{Pool}_\mathrm{3D}^{i^*}$. To convert the library-derived 3D pose into metric units, a scaling factor based on the human's predefined height and the observed skeletal length in pixels is applied. First, compute the current pixel height $h_{\mathrm{pixel}}$ of the skeleton as:
\begin{equation}
h_{\text{pixel}} = \max(\mathbf{P}_\mathrm{2D}(:,2)) - \min(\mathbf{P}_\mathrm{2D}(:,2)).
\end{equation}\par
Given the human's real-world height $h_{\mathrm{real}}$ in meters, the primary scaling factor is $\alpha_\mathrm{hum} = \frac{h_{\mathrm{real}}}{h_{\mathrm{pixel}}}$. This factor scales the 2D keypoints in the image plane (X,Z-axes):
\begin{equation}
\mathbf{P}_{\mathrm{meters, orig, xz}} = \mathbf{P}_\mathrm{2D} \cdot \alpha_\mathrm{hum}.
\end{equation}\par
For the depth dimension (Y-axis), a reduced scaling is employed to maintain realistic body proportions. Human body depth is typically a fraction of height, thus a depth ratio $\beta_\mathrm{hum} = 0.2$ is introduced, assuming depth is approximately $20\%$ of height. The depth $\mathbf{P}_\mathrm{3D}(:,3)$ is first centered around their mean $\overline{\mathbf{P}_\mathrm{3D}(:,3)}$ to avoid offsets:
\begin{equation}
\mathbf{P}_\mathrm{3D,depth,centered} = \mathbf{P}_\mathrm{3D}(:,3) - \overline{\mathbf{P}_\mathrm{3D}(:,3)}.
\end{equation}\par
The scaled depth is then:
\begin{equation}
\mathbf{P}_\mathrm{meters,orig,y} = \mathbf{P}_\mathrm{3D,depth,centered} \cdot \alpha_\mathrm{hum} \cdot \beta_\mathrm{hum}.
\end{equation}\par
The initial metric pose combines these components:
\begin{equation}
\mathbf{P}_{\mathrm{meters, orig}} = [\mathbf{P}_{\mathrm{meters, orig, xz}}, \mathbf{P}_{\mathrm{meters, orig, y}}].
\end{equation}\par
\begin{figure}
    \centering
    \includegraphics[width=0.48\textwidth]{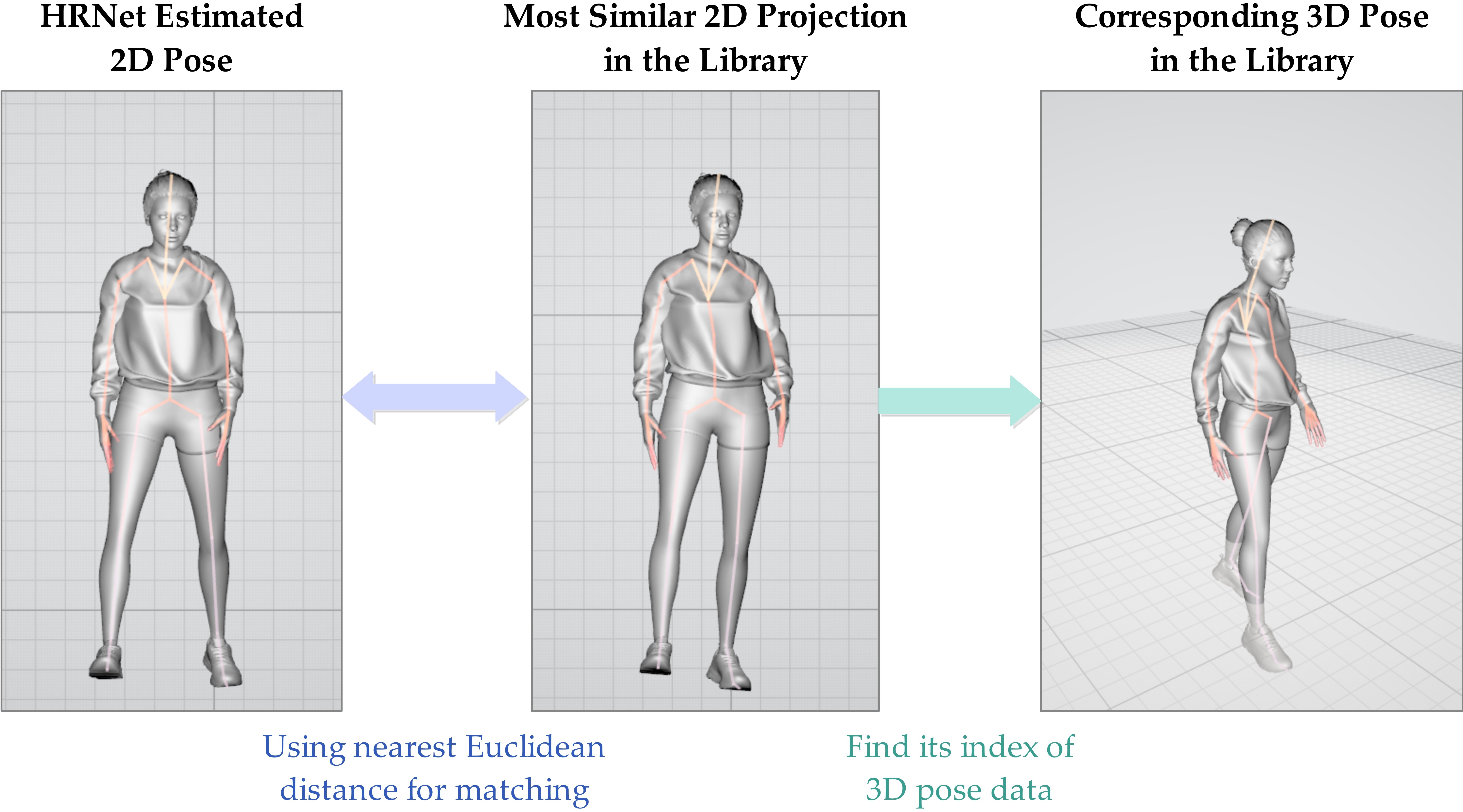}
    \caption{Schematic of the propsoed 2D to 3D pose matching method.}
    \label{Matching}
    \vspace{-0.3cm}
\end{figure}\par
To anchor the pose in a global metric coordinate system, the absolute position using the hip center as a reference is tracked. The hip center in the original metric pose is:
\begin{equation}
\begin{aligned}
\mathbf{Hip}_{\text{orig}} = &\left[ \overline{\mathbf{P}_{\text{meters, orig}}([9,12],1)}, \overline{\mathbf{P}_{\text{meters, orig}}([9,12],2)}, \right.\\&\left. \overline{\mathbf{P}_{\text{meters, orig}}([9,12],3)} \right]
\end{aligned}.
\end{equation}\par
The relative pose is computed by subtracting this center:
\begin{equation}
\begin{gathered}
\mathbf{P}_x^{\mathrm{rel}} = \mathbf{P}_{\mathrm{meters, orig}}(:,1) - \mathbf{Hip}_{\mathrm{orig}}(1)\\
\mathbf{P}_y^{\mathrm{rel}} = \mathbf{P}_{\mathrm{meters, orig}}(:,3) - \mathbf{Hip}_{\mathrm{orig}}(3)\\
\mathbf{P}_z^{\mathrm{rel}} = -\left(\mathbf{P}_{\mathrm{meters, orig}}(:,2) - \mathbf{Hip}_{\mathrm{orig}}(2)\right)
\end{gathered}.
\end{equation}\par
The Z-coordinate inversion aligns the coordinate system with radar conventions. To ground the pose, an offset is applied based on the lowest ankle joint:
\begin{equation}
\begin{gathered}
z_{\mathrm{offset}} = \min\left(\mathbf{P}_z^{\mathrm{rel}}(11), \mathbf{P}_z^{\mathrm{rel}}(14)\right)\\
\mathbf{P}_z^{\mathrm{grounded}} = \mathbf{P}_z^{\mathrm{rel}} - z_{\mathrm{offset}}
\end{gathered}.
\end{equation}\par
The absolute position $\mathbf{Pos}_{\mathrm{abs}} = (x_{\mathrm{abs}}, y_{\mathrm{abs}}, z_{\mathrm{abs}})$ is updated frame-by-frame via pixel displacement of the hip center, scaled by $\alpha_\mathrm{hum}$, with $z_{\mathrm{abs}} = 0$ initially, where $x_{\mathrm{abs}}, y_{\mathrm{abs}}, z_{\mathrm{abs}}$ are the absolute coordinates. The final metric pose is:
\begin{equation}
\mathbf{P}_{\mathrm{meters}} = [\mathbf{P}_x^{\mathrm{rel}} + x_{\mathrm{abs}}, \mathbf{P}_y^{\mathrm{rel}} + y_{\mathrm{abs}}, \mathbf{P}_z^{\mathrm{grounded}} + z_{\mathrm{abs}}]
\end{equation}.\par
The resulted $\mathbf{P}_{\mathrm{meters}}$ will be used for subsequent Kalman filtering-based temporal smoothing.\par

\subsection{Temporal Smoothing for 3D Pose}
To mitigate jitter, a Kalman filter is applied to each joint's 3D trajectory \cite{KalmanFiltering}. The temporal relationships between joints for prediction and constraint is used, resulting in smoother estimated 3D poses. Assume the state vector of the estimated 3D pose is:
\begin{equation}
\mathbf{st}_{\mathrm{meters}}=[\mathbf{P}_{\mathrm{meters}}, \dot{\mathbf{P}}_{\mathrm{meters}}]^{\top},
\end{equation}
with transition matrix:
\begin{equation}
\mathbf{TransP} = \begin{bmatrix}
\mathbf{I}_{3\times 3} & \Delta t_f \cdot \mathbf{I}_{3\times 3}\\
\mathbf{0}_{3\times 3} & \mathbf{I}_{3\times 3}
\end{bmatrix}.
\end{equation}\par
The measurement matrix is $\mathbf{MeasP}=[\mathbf{I}_{3\times 3},\mathbf{0}_{3\times 3}]$. Process and measurement noise covariances are set empirically. For each joint, prediction and correction steps yield a smoothed pose, enhancing stability during occlusions or detection losses.\par
The above pipeline enables accurate 3D kinematics for downstream radar simulation, bridging computer vision outputs with physics-based modeling.\par
\begin{figure*}
    \centering
    \includegraphics[width=\textwidth]{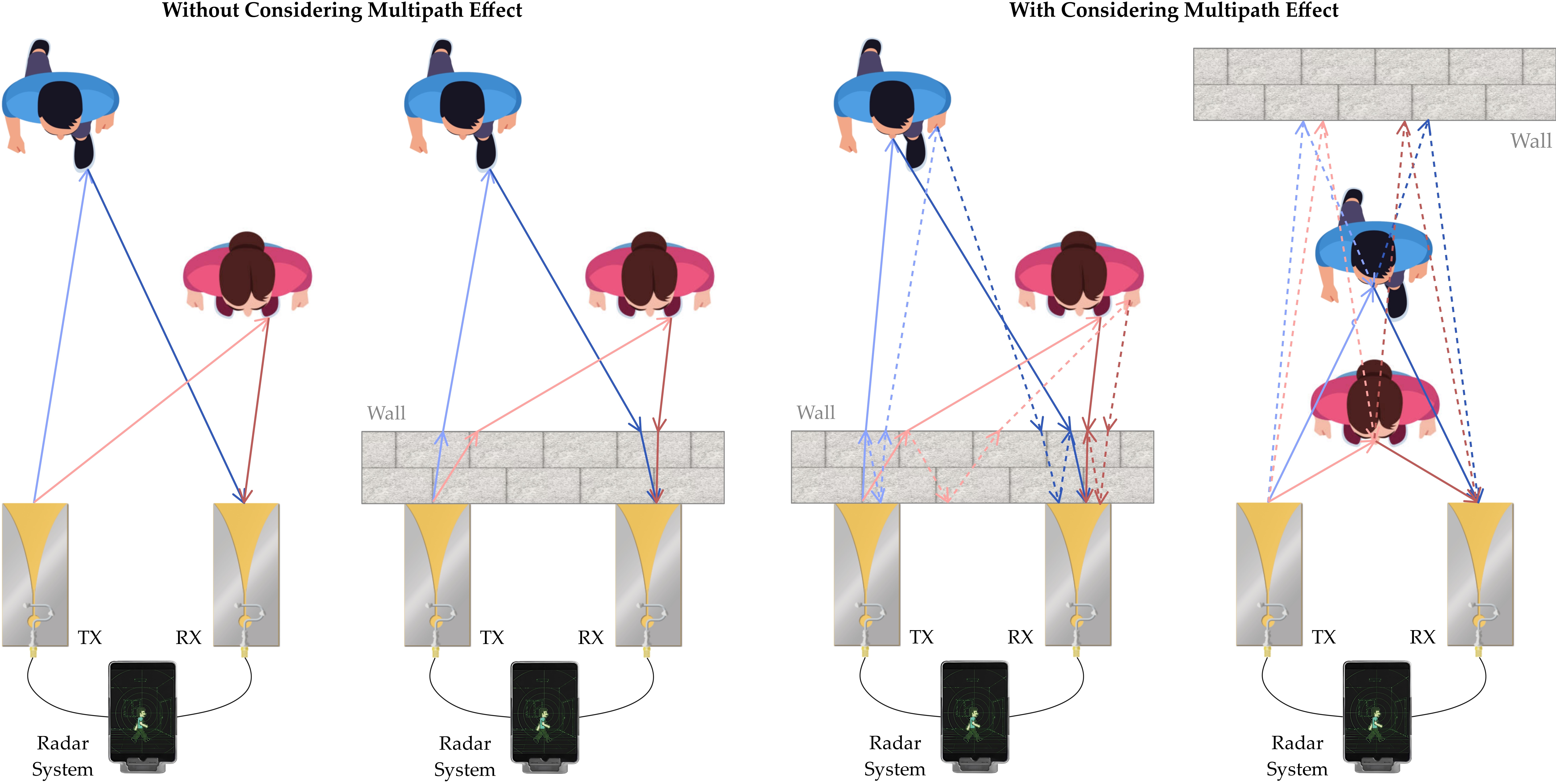}
    \caption{Echo modeling of the radar simulation. Schematics from left to right represent: Free-space, TTW, TTW with multipath, free-space with multipath.}
    \label{Echo_Model}
    \vspace{-0.1cm}
\end{figure*}\par

\section{Radar Module}
In this section, pose interpolation, filtering, and echo modeling in free-space and TTW scenarios with multipath effects are first introduced. Pulse compression and MTI are then employed to generate the RTM, while the STFT is used to generate the DTM. Next, the DnCNN model is utilized to perform denoising and micro-Doppler signature enhancement, and the ridge features on DTM are extracted using MLE.\par

\subsection{Pose Interpolation and Filtering}
Assume the temporally smoothed 3D human pose is denoted as $\mathbf{P}_{j,p}$, where $p\in \mathbb{Z}^{+}\cap [1,P]$ represents the $p^\mathrm{th}$ human target detected, and $j \in \mathbb{Z}^{+}\cap [1, K_\mathrm{HRNet}]$ represents the $j^\mathrm{th}$ joint. To convert the video timestamp $t_f$ to the radar pulse $t_m$, linear interpolation is applied to the 3D pose results \cite{LinearInterpolation}:
\begin{equation}
\begin{gathered}
\mathbf{P'}_{j,p}(t_m) = \mathbf{P}_{j,p}(t_f) + \frac{\mathbf{P}_{j,p}(t_f+\Delta t_f) - \mathbf{P}_{j,p}(t_f)}{\Delta t_f}(t_m - t_f) \\ t_f \le t_m < t_f+\Delta t_f
\end{gathered}.
\end{equation}\par
The resulted 3D pose $\mathbf{P'}_{j,p}(t_m)$ is processed under Savitzky-Golay filtering to remove the noise and outliers. Let $\mathbf{x}_\mathrm{Sgolay}$ represent a single coordinate's time series from $\mathbf{P'}_{j,p}$. For a point at time $t_m$, a window of $N_\mathrm{Sgolay}=2m+1$ points is considered, indexed by a local variable $n \in \{-m, -m+1,\ldots,0,\ldots, m-1, m\}$. The data points in this window are $\mathbf{x}_\mathrm{Sgolay}(t_{m+n})$. A polynomial $p_\mathrm{Sgolay}(n)$ is fitted to these points:
\begin{equation}
p_\mathrm{Sgolay}(n)=\sum_{k=0}^{d_\mathrm{Sgolay}}c_{\mathrm{Sgolay},k} \cdot n^{k},
\end{equation}
where $d_\mathrm{Sgolay}$ is the order of Savitzky-Golay filtering \cite{Savitzky-Golay}, the coefficients $c_{\mathrm{Sgolay},k}$ are found by minimizing the sum of squared errors $E_\mathrm{Sgolay}$ between the polynomial and the data:
\begin{equation}
\begin{aligned}
E_\mathrm{Sgolay} &= \sum_{n=-m}^{m} \left( p_\mathrm{Sgolay}(n) - \mathbf{x}_\mathrm{Sgolay}(t_{m+n}) \right)^2 \\&= \sum_{n=-m}^{m} \left( \left(\sum_{k=0}^{d_\mathrm{Sgolay}} c_{\mathrm{Sgolay},k} \cdot n^k\right) - \mathbf{x}_\mathrm{Sgolay}(t_{m+n}) \right)^2
\end{aligned}.
\end{equation}\par
Solve the minimization problem of $E_\mathrm{Sgolay}$ using the least squares method to obtain the solution for $c_{\mathrm{Sgolay},k}$. Substitute the smoothing polynomial $p_\mathrm{Sgolay}$ for $\mathbf{x}_\mathrm{Sgolay}$ to reconstruct the twice-smoothed 3D pose matrix $\mathbf{P''}_{j,p}$, which serves as the input for the subsequent echo simulation.\par
 
\subsection{Echo Modeling}
The linear frequency-modulated (LFM) wave is used in this work \cite{LFM}. The instantaneous frequency of LFM is defined as $f(t)=f_c+\mu t$, where $f_c$ is the carrier frequency, $\mu$ is the frequency-modulated slope, $t$ is the fast time. The phase of the signal is the integral of the angular frequency $\omega(t)=2\pi f(t)$:
\begin{equation}
\theta_\mathrm{tx}(t) = \int_0^t 2\pi(f_c + \mu\xi)d\xi = 2\pi\left(f_c t + \frac{1}{2}\mu t^2\right).
\end{equation}\par
The complex transmitted signal is therefore represented as:
\begin{equation}
s_\mathrm{tx}(t) = \exp\left(j\theta_\mathrm{tx}(t)\right) = \exp\left(j2\pi\left(f_c t + \frac{1}{2}\mu t^2\right)\right).
\end{equation}\par
The echo from a scatterer at a distance corresponding to a time delay $\tau$ is received as $s_\mathrm{rx}(t) = A\cdot s_\mathrm{tx}(t - \tau)$ with its phase representing as:
\begin{equation}
\begin{aligned}
\theta_{rx}(t) &= 2\pi\left(f_c(t-\tau) + \frac{1}{2}\mu(t-\tau)^2\right) \\&= 2\pi\left(f_c t - f_c\tau + \frac{1}{2}\mu t^2 - \mu t\tau + \frac{1}{2}\mu\tau^2\right)
\end{aligned}.
\end{equation}\par
This received signal is mixed with the transmitted signal and low-pass filtered known as de-chirping-based pulse compression, which multiplies $s_\mathrm{rx}(t)$ by the complex conjugate of $s_\mathrm{tx}(t)$, which is $s^{*}_\mathrm{tx}(t)=\exp (-j\theta_\mathrm{tx}(t))$. The phase of the resulting intermediate frequency (IF) signal is the difference between the received and transmitted phases:
\begin{equation}
\begin{aligned}
\theta_\mathrm{if}(t) &= \theta_\mathrm{rx}(t) - \theta_\mathrm{tx}(t) \\&= 2\pi\left[\left(f_c t - f_c\tau + \frac{1}{2}\mu t^2 - \mu t\tau + \frac{1}{2}\mu\tau^2\right) \right.\\& \left.\quad - \left(f_c t + \frac{1}{2}\mu t^2\right)\right]\\
\Rightarrow \quad \theta_\mathrm{if}(t) & = 2\pi\left( - f_c\tau - \mu t\tau + \frac{1}{2}\mu\tau^2 \right)
\end{aligned}.
\end{equation}\par
By convention and noting that the sign of the frequency term depends on the specific de-chirping architecture, the phase is typically expressed to yield a positive beat frequency. The resulting complex baseband signal for a single scatterer is:
\begin{equation}
s_\mathrm{if}(t, \tau) = A \cdot \exp\left(j2\pi\left(f_c\tau + \mu\tau t - \frac{1}{2}\mu\tau^2\right)\right).
\label{Sif}
\end{equation}\par
When a plane wave encounters a dielectric wall, its propagation is governed by the material's properties \cite{Multipath1}. Starting from Maxwell's curl equations in a lossy medium, the Helmholtz wave equation is derived:
\begin{equation}
\nabla^2\mathbf{E} - \gamma^2\mathbf{E} = 0,
\end{equation}
where $\gamma$ is the complex propagation constant. For a medium with complex permittivity $\epsilon_c=\epsilon' - j\epsilon'' = \epsilon_r\epsilon_0(1-j\tan \delta)$, where $\tan \delta$ is the loss tangent, and permeability $\mu_0$, $\gamma$ can be given by: 
\begin{equation}
\gamma = \sqrt{j\omega\mu_0(\sigma + j\omega\epsilon')} = j\omega\sqrt{\mu_0\epsilon_c} = \alpha + j\beta,
\end{equation}
where $\alpha$ is the attenuation constant and $\beta$ is the phase constant. The intrinsic impedance of the wall $\eta_\mathrm{wall}$ is:
\begin{equation}
\eta_\mathrm{wall} = \sqrt{\frac{j\omega\mu_0}{\sigma + j\omega\epsilon'}} = \sqrt{\frac{\mu_0}{\epsilon_c}}.
\end{equation}\par
At an air-wall interface (Assuming from medium 1 to medium 2), boundary conditions require the tangential components of $E^w$ and $H^w$ to be continuous. For a normally incident wave, this leads to:
\begin{equation}
E^w_{i} + E^w_{r} = E^w_{t},\quad H^w_{i} + H^w_{r} = H^w_{t}.
\end{equation}\par
Using the relation $H^w = E^w/\eta$, where $\eta_1=\eta_0$ as the impedance of air and $\eta_2=\eta_\mathrm{wall}$:
\begin{equation}
\frac{E^w_{i}}{\eta_1} - \frac{E^w_{r}}{\eta_1} = \frac{E^w_{t}}{\eta_2}.
\end{equation}\par
Solving these two simultaneous equations for the reflection coefficient $\Gamma^w = E^w_r/E^w_i$ and transmission coefficient $T^w = E^w_t/E^w_i$ yields:
\begin{equation}
\begin{gathered}
\Gamma^w_{12} = \frac{\eta_2 - \eta_1}{\eta_2 + \eta_1} = \frac{\eta_\mathrm{wall} - \eta_0}{\eta_\mathrm{wall} + \eta_0}\\
T^w_{12} = 1 + \Gamma^w_{12} = \frac{2\eta_2}{\eta_2 + \eta_1} = \frac{2\eta_\mathrm{wall}}{\eta_\mathrm{wall} + \eta_0}
\end{gathered}.
\end{equation}\par
For the wall-air interface (Assuming from medium 2 to medium 1), the coefficients are $\Gamma^w_{21} = (\eta_1-\eta_2)/(\eta_1+\eta_2) = -\Gamma^w_{12}$ and $T^2_{21} = 1 + \Gamma^w_{21}$.\par
When the wall is present, multipath effects should also be considered. The total IF signal is the coherent sum of signals from all propagation paths \cite{Multipath2}:\par
\textbf{(1) Direct Transmission of Target Behind the Wall:}\par
The signal passes TTW with thickness $d_w$, reflects off the target, and passes back TTW. The one-way complex transmission factor is the product of transmission at the first interface, propagation TTW, and transmission at the second interface: $T^w_{12}\cdot e^{-\gamma d_w} \cdot T^w_{21}$. The total two-way complex transmission factor $T^w_\mathrm{tw}$ is the square of the one-way factor:
\begin{equation}
T^w_\mathrm{tw} = \left( T^w_{12}\cdot T^w_{21} \cdot e^{-\gamma d_w} \right)^2.
\end{equation}\par
The propagation time inside the wall is longer than in air. The phase velocity in the wall is $v_p = \frac{\omega}{\beta} = \frac{c}{n_{\mathrm{wall}}}$, where $n_{\mathrm{wall}} = \operatorname{Re}\left\{\sqrt{\frac{\epsilon_c}{\epsilon_0}}\right\}$ is the refractive index. The extra time delay for a one-way trip TTW is $\Delta t = \frac{d_w}{v_p} - \frac{d_w}{c} = \frac{d_w(n_{\mathrm{wall}} - 1)}{c}$. For a two-way path, the total extra delay is $2\Delta t$. The effective time delay, $\tau'$, is:
\begin{equation}
\tau'_{j,p}(t_m) = \frac{R_{j,p}(t_m)}{c} + \frac{2d_w(n_{\mathrm{wall}} - 1)}{c},
\end{equation}
where $R_{j,p}(t_m)$  is the range of the $p^\mathrm{th}$ target and its $j^\mathrm{th}$ joint. The IF signal for this path is found by using $\tau'$ in Eq. (\ref{Sif}) and scaling by $T^w_\mathrm{tw}$.\par
\textbf{(2) Internal Multipath of Target Behind the Wall:}\par
A multipath component arises from internal reverberations. For a signal that bounces once inside the wall on each joint, the following paths should be considered: Enters wall ($T^w_{12}$), propagates to back face ($e^{-\gamma d_w}$) $\rightarrow$ Reflects from back face ($\Gamma^w_{21}$), propagates to front face ($e^{-\gamma d_w}$) $\rightarrow$ Reflects from front face ($\Gamma^w_{21}$), propagates to back face ($e^{-\gamma d_w}$) $\rightarrow$ Exits wall ($T^w_{21}$). Therefore, the one-way factor is $T^w_{12} \cdot {\Gamma^w_{21}}^{2} \cdot T^w_{21} \cdot e^{-3\gamma d_w}$. The two-way complex factor, $T_{\mathrm{multipath}}$, is the square of this:
\begin{equation}
T_{\text{multipath}} = \left( T^w_{12} \cdot {\Gamma^w_{21}}^{2} \cdot T^w_{21} \cdot e^{-3\gamma d_w} \right)^{2}.
\end{equation}\par
This path travels an additional distance of $2d_w$ within the wall for each joint compared to the direct path. The effective two-way time delay $\tau''$ is:
\begin{equation}
\tau''_{j,p}(t_m) = \tau'_{j,p}(t_m) + \frac{4d_w \cdot n_{\mathrm{wall}}}{c}.
\end{equation}\par
\textbf{(3) Reflection Multipath of Target in Front of the Wall:}\par
Using the method of images, where virtual transmitter $\mathbf{P}'_{\mathrm{tx}}$ and receiver $\mathbf{P}'_{\mathrm{rx}}$ are reflections of the true antennas across the wall face, the multipath path lengths are calculated geometrically. The corresponding echoes are scaled by the reflection coefficient $\Gamma^w_{12}$ for each bounce off the wall. The time delays ($\tau_a$, $\tau_b$, $\tau_c$) for the three primary multipath components are:
\begin{equation}
\begin{gathered}
\tau_a = \frac{\| \mathbf{P}''_{j,p} - \mathbf{P}'_{\mathrm{tx}} \| + \| \mathbf{P}''_{j,p} - \mathbf{P}'_{\mathrm{rx}} \|}{c}\\
\tau_b = \frac{\| \mathbf{P}''_{j,p} - \mathbf{P}'_{\mathrm{tx}} \| + \| \mathbf{P}''_{j,p} - \mathbf{P}'_{\mathrm{rx}} \|}{c}\\
\tau_c = \frac{\| \mathbf{P}''_{j,p} - \mathbf{P}'_{\mathrm{tx}} \| + \| \mathbf{P}''_{j,p} - \mathbf{P}'_{\mathrm{rx}} \|}{c}
\end{gathered}.
\end{equation}\par
\begin{figure}
    \centering
    \includegraphics[width=0.48\textwidth]{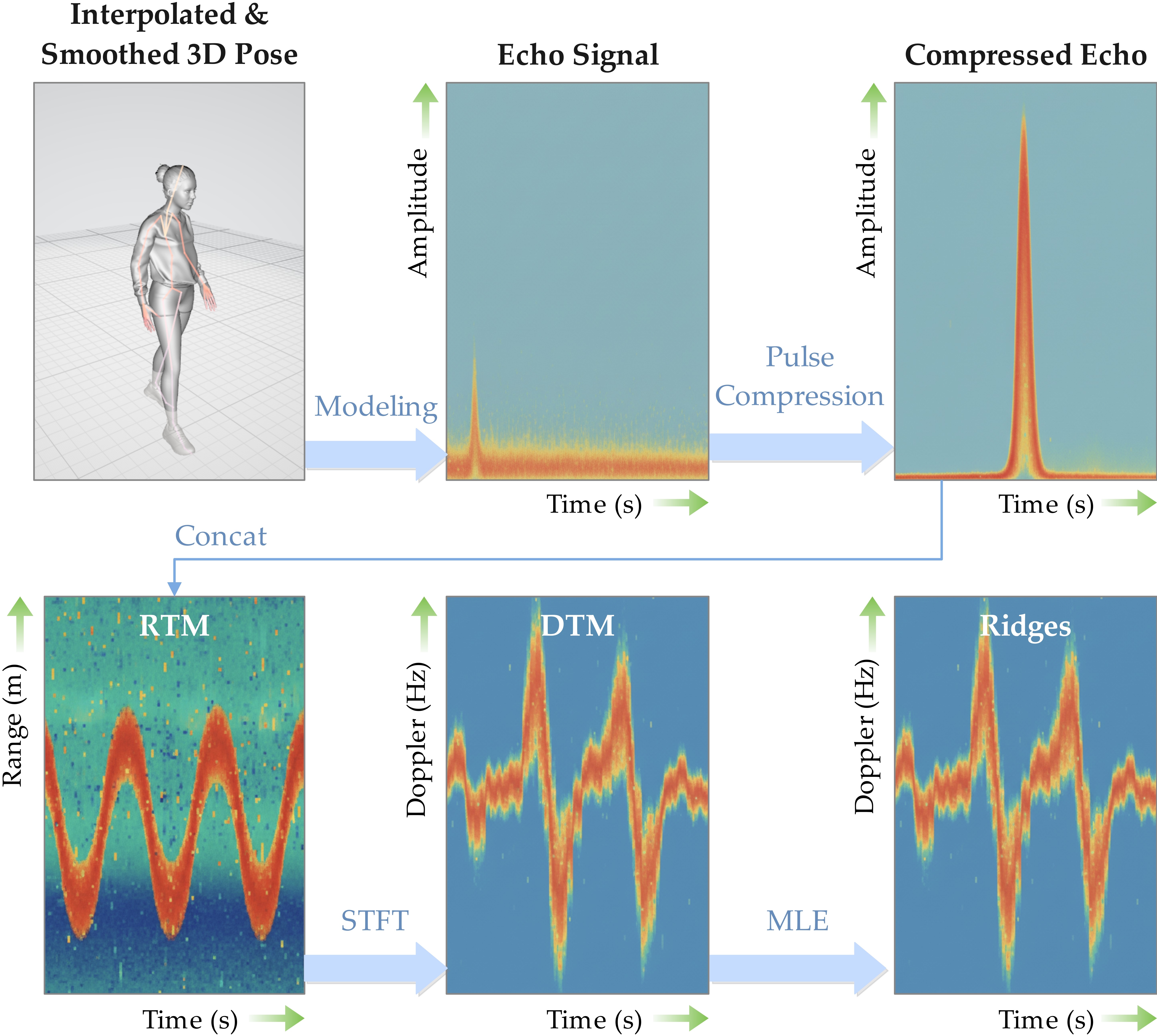}
    \caption{The flowchart of generating R/DTM with ridge extraction.}
    \label{RDTM_Generation}
    \vspace{-0.2cm}
\end{figure}\par
The final IF signal, $S_{\mathrm{if}}(t, t_m)$, is the coherent sum of the primary path and all applicable multipath components, plus complex white Gaussian noise $\mathbf{n}(t)$.\par

\subsection{RTM and DTM Generation}
As shown in Fig. \ref{RDTM_Generation}, the RTM is generated by applying the fast Fourier transform (FFT) along the fast-time axis of the total IF signal for each pulse \cite{RDTM}:
\begin{equation}
\mathrm{RTM}(r, t_m) = \left| \mathcal{F} \left\{ S_{\text{if}}(t, t_m) \right\} \right|,
\end{equation}
where $r$ is the range bin of the RTM. A two-pulse canceller is applied to the complex range profiles to suppress static clutter:
\begin{equation}
\mathrm{RTM_{MTI}}(r, t_m) = \left| \mathcal{F} \left\{ S_{\text{if}}(t, t_m) \right\} - \mathcal{F} \left\{ S_{\text{if}}(t, t_{m-1}) \right\} \right|.
\end{equation}\par
The DnCNN model is proposed for noise suppression and micro-Doppler signature enhancement on MTI-processed RTM \cite{DnCNN}. The DnCNN model takes the noisy map as input, preprocessing them using a $3\times 3$ convolutional layer with $64$ channels and a rectified linear unit (ReLU) activation function. The model then stacks bottlenecks consisting of $18$ repetitions of $3\times 3$ convolutions, batch normalization, and ReLU activation functions in sequence. Finally, a $3 \times 3$ convolution with a channel count of $1$ is used to estimate the noise mask map. Subtract the input image from the noise mask to obtain a map that is denoised and feature-enhanced.\par
The primary objective of the loss function of DnCNN model is to minimize the mean squared error between the residual map output by the network and the true label map, expressed as:
\begin{equation}
\mathcal{L}_\mathrm{DnC}(\theta_\mathrm{DnC}) = \frac{\left\| \mathcal{Q}(y_{\mathrm{DnC},i}; \theta_\mathrm{DnC}) - (y_{\mathrm{DnC},i} - x_{\mathrm{DnC},i}) \right\|^2}{2N_\mathrm{DnC}}, 
\end{equation}
where $y_{\mathrm{DnC},i}$ is the noisy input map, $x_{\mathrm{DnC},i}$ is the true label map, $\mathcal{Q}(y_{\mathrm{DnC},i})$ is the inference output of the DnCNN, $N_\mathrm{DnC}$ is the number of training batch, and $\theta_\mathrm{DnC}$ is the parameter of the DnCNN model. Thus, it is important to note that the input to the DnCNN must be a map with absolute values. Furthermore, the input for subsequent DTM generation must be the MTI-processed RTM prior to DnCNN enhancement.\par
The DTM is generated via the STFT of a slow-time signal, which is derived from the range summation of the MTI-processed RTM:
\begin{equation}
\begin{aligned}
\mathrm{DTM}(f_d, \tau) &= \left| \int_{-\infty}^{\infty} \sum_r(\mathrm{RTM_{MTI}}(r, t_m)) \right.\\&\left. \quad \cdot \mathrm{win}(t_m - \tau) e^{-j 2 \pi f_d t_m} \, dt_m \right|
\end{aligned},
\end{equation}
where $f_d$ is the Doppler axis of DTM, and $\mathrm{win}(\cdot)$ is the window function of STFT, with the preset window length and overlap ratio. DTM can also undergo denoising and micro-Doppler signature enhancement via DnCNN.\par

\subsection{Ridge Feature Extraction}
The most prominent micro-Doppler signature is extracted from the DTM by identifying the $N_r$ time-frequency ridges \cite{TFRidges}, $\{f_{\mathrm{ridge},k}(\tau) \mid k = 1,\dots,N_r\}$, corresponding to the paths of MLE. This is achieved by finding the $N_r$ largest local maxima of $\mathrm{DTM}(f_d, \tau)$ for each time instant $\tau$ and connecting them through time:
\begin{equation}
\{ f_{\text{ridge},k}(\tau) \}_{k=1}^{N_r} = {\arg \max_{f_d}} \left\{ \mathrm{DTM}(f_d, \tau) \right\}|_{N_r}.
\end{equation}\par
\begin{figure}
    \centering
    \includegraphics[width=0.48\textwidth]{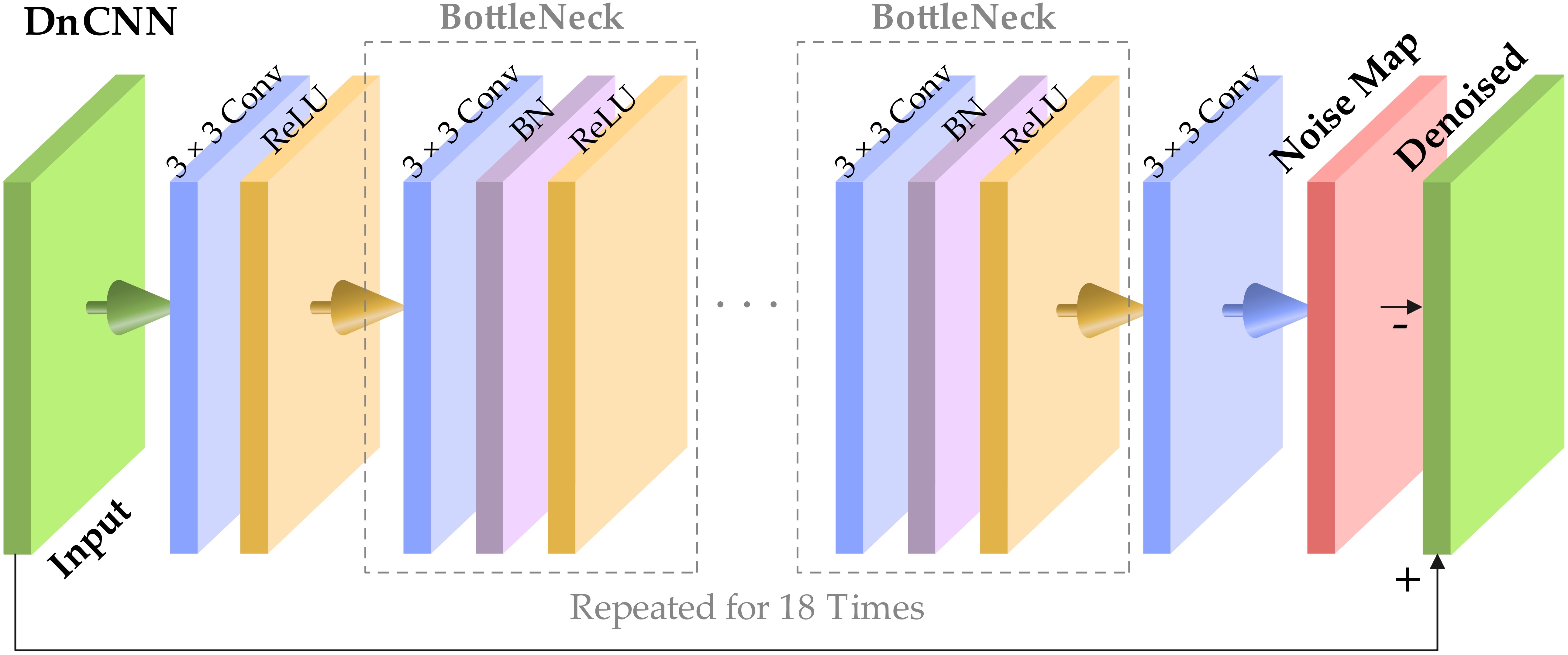}
    \caption{Structure design of the DnCNN model.}
    \label{DnCNN}
    \vspace{-0.3cm}
\end{figure}\par
The generated DTM images, the sequence of extracted time-frequency ridges, or the curve images plotted from time-frequency ridges can all be fed into neural network models to achieve HAR \cite{HARBasis}.\par
\begin{figure*}
    \centering
    \includegraphics[width=\textwidth]{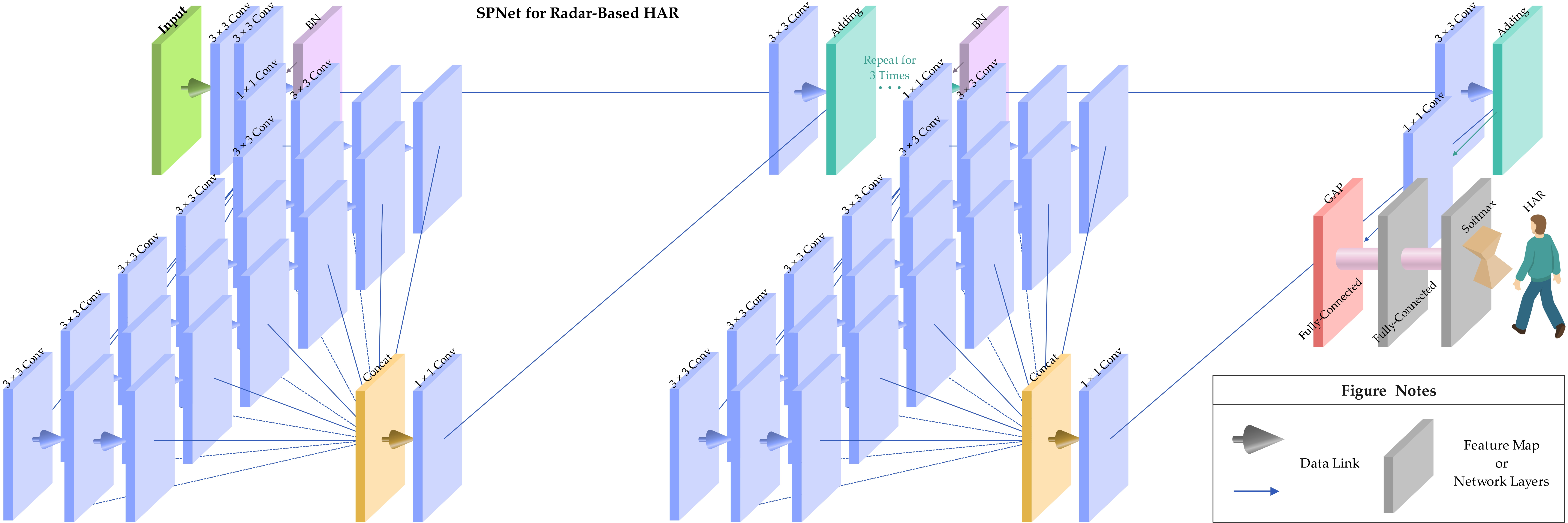}
    \caption{Structure for the designed SPNet used for radar-based HAR task.}
    \label{SPNet}
    \vspace{-0.3cm}
\end{figure*}\par
\begin{table}[!ht]
\begin{center}
\caption{Experimental Parameter Settings.\label{tab:SimParams}}
\vspace{-0.0cm}
\resizebox{0.48\textwidth}{!}{
\begin{tabular}{ccc}
\hline\hline
\textbf{Parameter} & \textbf{Free-Space} & \textbf{Through-the-Wall} \\
\hline
\multicolumn{3}{c}{\textbf{Radar Parameters}} \\
\hline
Carrier Frequency & $77\mathrm{~GHz}$ & $2\mathrm{~GHz}$ \\
Bandwidth & $4\mathrm{~GHz}$ & $1\mathrm{~GHz}$ \\
Pulse Repetition Frequency & $8192\mathrm{~Hz}$ & $128\mathrm{~Hz}$ \\
Sampling Frequency & \multicolumn{2}{c}{$10\mathrm{~MHz}$} \\
Pulse Duration & \multicolumn{2}{c}{$40~\mu s$} \\
Transmitter Position & \multicolumn{2}{c}{$(-0.1, 0, 1.5)~m$} \\
Receiver Position & \multicolumn{2}{c}{$(0.1, 0, 1.5)~m$} \\
Antenna Gain & \multicolumn{2}{c}{$10\mathrm{~dBi}$} \\
Antenna Isolation & \multicolumn{2}{c}{$20\mathrm{~dB}$} \\
SNR & \multicolumn{2}{c}{$50\mathrm{~dB}$} \\
\hline
\multicolumn{3}{c}{\textbf{Wall Parameters}} \\
\hline
Wall Center Position & / & $(-0.1,0,1.5)~m$ \\
Wall Dimensions  & / & $(5, 0.24, 2.5)~m$ \\
Relative Dielectric Constant  & / & $6$ \\
Loss Tangent & / & $0.03$ \\
\hline\hline
\end{tabular}
}
\end{center}
\vspace{-0.2cm}
\end{table}\par
\begin{table}[!ht]
\begin{center}
\caption{Uniform Hyperparameters for Networks.\label{Training Settings}}
\vspace{-0.1cm}
\resizebox{0.48\textwidth}{!}{
\begin{tabular}{cc}
\hline\hline
\textbf{Name of Hyperparameters}             & \textbf{Value}          \\ \hline
Batch Size                      & $64$                     \\
Total Epoches                   & $20$                    \\
Initial Learning Rate$^{1}$     & $0.00147$                   \\
Regularization Method             & $L - 2$           \\
Optimizer                       & Adam  \\
Solidified Model                & Best Epoch                \\
Training Dataset                & $2880$                     \\
Validation Dataset              & $720$                     \\
Hardware of Training and Validation   & NVIDIA RTX 3060 OC \\
Software of Training and Validation  & Matlab R2025b   \\
\hline\hline
\end{tabular}
}
\end{center}
\vspace{-0.4cm}
\end{table}\par

\section{Hybrid Parallel-Serial Neural Network Model for Radar-Based HAR}
As shown in Fig. \ref{SPNet}, in this paper, the SPNet is proposed as an example for achieving the mapping from R/DTM to activity labels \cite{SPGNet}. The input data of SPNet represented as a multi-dimensional tensor is first fed into an initial convolutional block consisting of a $3\times 3$ convolution layer followed by batch normalization (BN). This is succeeded by a series of densely connected convolutional layers, where each subsequent $3\times 3$ convolution layer receives concatenated feature maps from all preceding layers for promoting feature reuse. A $1\times 1$ convolution is applied after concatenation to reduce dimensionality and fuse features efficiently. Following the dense block, an adding operation is performed to merge residual connections, and this structure is repeated three times to form a multi-stage feature extraction pipeline. In each repetition, additional $3\times 3$ convolution layers with batch normalization are incorporated, and $1\times 1$ convolutions are utilized for channel adjustment. The output from these repeated blocks is then directed to a final convolutional stage, which includes further $3\times 3$ convolutions and concatenation, leading to a global average pooling (GAP) layer that aggregates spatial features into a compact representation. The pooled features are subsequently passed through a fully-connected layer to produce class logits, after which a softmax activation is applied to generate probability distributions for HAR categories.\par
The cross-entropy loss function for the network is designed to measure the discrepancy between predicted probabilities and true activity labels in a multi-class classification setting. Categorical cross-entropy is employed for SPNet, where the loss for a single sample is calculated as the negative sum of the true label distribution multiplied by the logarithm of the predicted probabilities:
\begin{equation}
\mathcal{L}_\mathrm{SPNet} = -\sum_{i=1}^{\mathrm{Cls}} y_{\mathrm{SPNet},i} \log(p_{\mathrm{SPNet},i}),
\end{equation}
where $\mathrm{Cls}$ represents the number of classes, $y_{\mathrm{SPNet},i}$ denotes the one-hot encoded true label for class $i$, and $p_{\mathrm{SPNet},i}$ signifies the predicted probability for class $i$. This loss is averaged over all samples in a batch during training to optimize the network parameters. Network models can be trained separately using RTM, DTM, channel-merged R/DTM, or DTM ridge images. New input data can be fed into the network for inference to obtain HAR results.\par

\section{Numerical Experiments}
A thorough validation of the data produced by the constructed simulato, including visualizations, comparative verifications, ablation experiments, and the suggested neural network's performance is provided and discussed detailedly in this section.\par
\begin{figure*}[!ht]
    \centering
    \includegraphics[width=\textwidth]{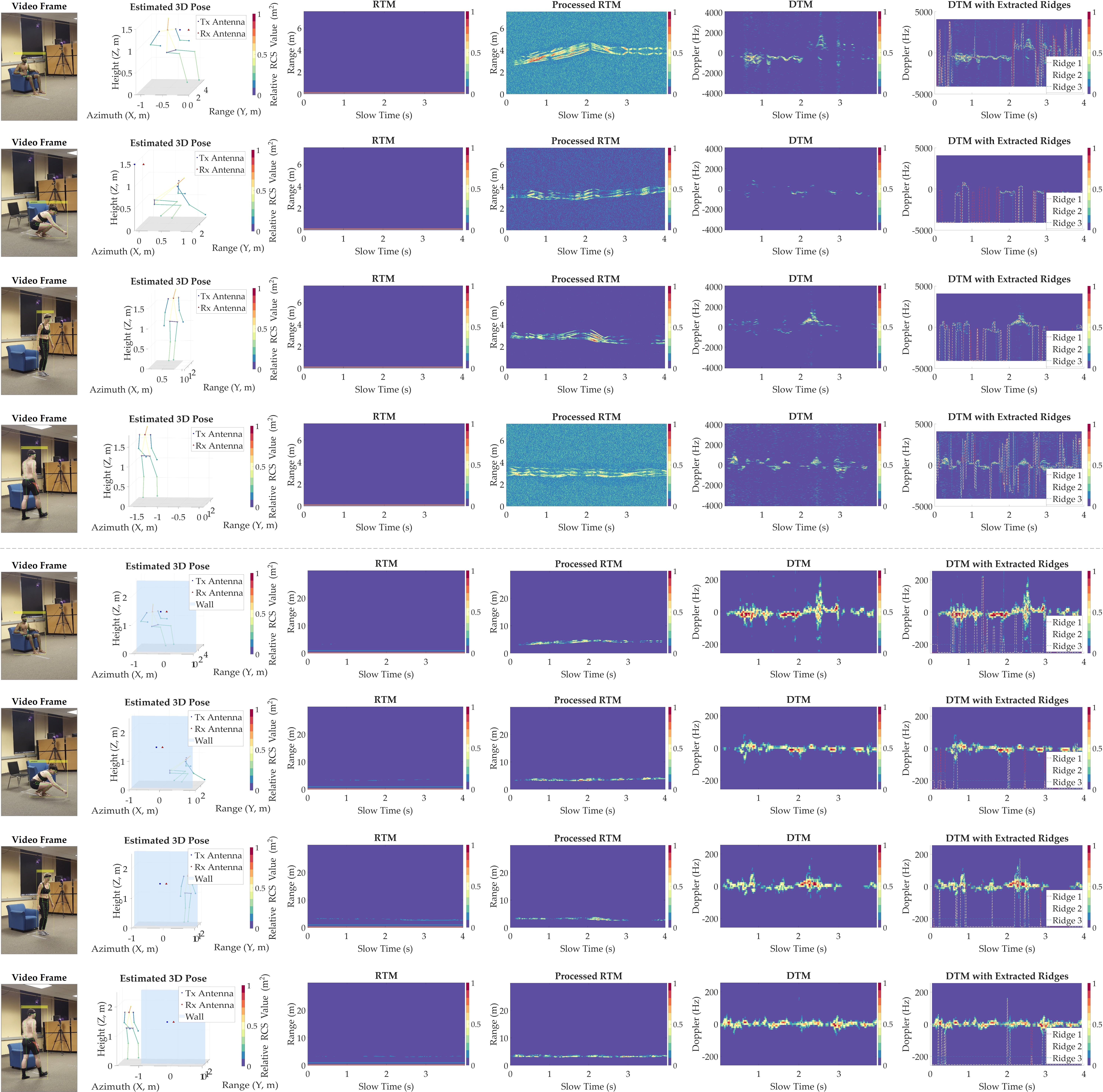}
    \caption{Visualizations under OSSet: The first row is for free-space sitting down activity, the second row is for grabbing, the third row is for standing up, the forth row is for walking, and row 5 to row 8 follow the same activities as the first four rows, with the simulation scenario being shifted to TTW.}
    \label{Visualizations_OpenSource}
    \vspace{-0.0cm}
\end{figure*}\par
\begin{figure*}[!ht]
    \centering
    \includegraphics[width=\textwidth]{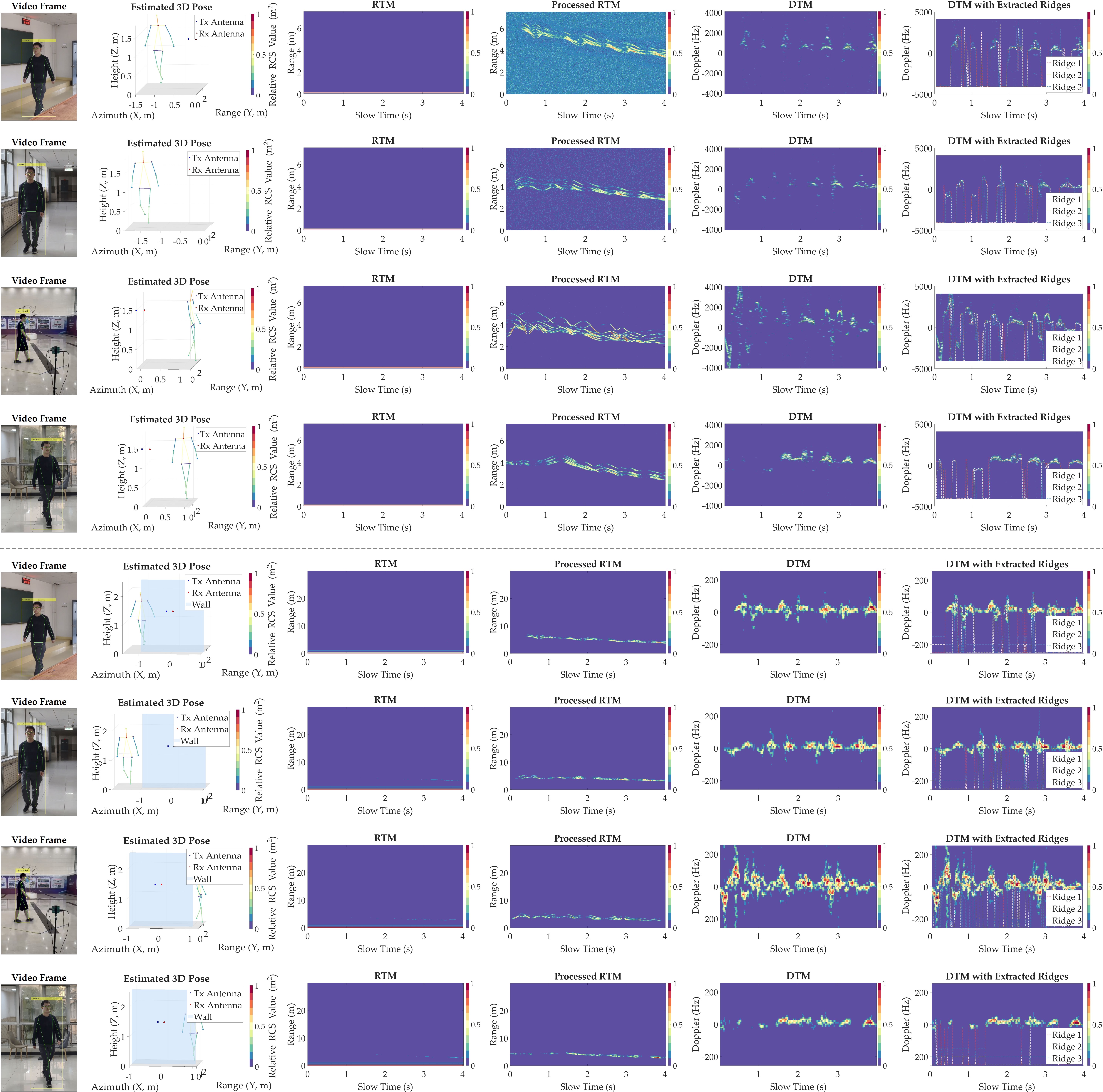}
    \caption{Visualizations using RWSet: Row 1 to 4 are used for free-space simulation taken from classroom, corridor, hall, and pantry, respectively, and row 5 to row 8 change the simulation scenario to TTW.}
    \label{Visualizations_RW}
    \vspace{-0.0cm}
\end{figure*}\par

\subsection{Experimental Settings}
The parameters of the kinematic model and the radar echo model can be changed in the simulator, as indicated in TABLE \ref{tab:SimParams}. Free-space detection \cite{Free-Space Detection} and TTW detection \cite{Through-the-Wall Detection} are the two scenarios used for analysis in this paper.\par
In free-space detection scenario, the radar's carrier frequency is set to $77\mathrm{~GHz}$ with a bandwidth of $4\mathrm{~GHz}$. A PRF of $8192\mathrm{~Hz}$ and a sampling frequency of $10\mathrm{~MHz}$ are used, with a pulse duration of $40~\mu s$. The transmitter and receiver are positioned at $(-0.1, 0, 1.5)~m$ and $(0.1, 0, 1.5)~m$, respectively. An antenna gain of $10\mathrm{~dBi}$ and an antenna isolation of $20\mathrm{~dB}$ are specified, and a signal-to-noise ratio (SNR) of $50\mathrm{~dB}$ is assumed.\par
In TTW detection scenario, the radar's carrier frequency is changed to $2\mathrm{~GHz}$ with a bandwidth of $1\mathrm{~GHz}$. A pulse repetition frequency of $128\mathrm{~Hz}$ is used. The center of the wall is placed at $(-0.1,0,1.5)~m$. The length, width, and height of the wall are $0.24$, $5$, and $2.5~m$, respectively. The relative dielectric constant and the loss tangent of the wall are $6$ and $0.03$, respectively. The rest of the parameters remain the same as free-space detection.\par
For experimental human activities, there are $12$ categories designed for the network training and inference labels, including: Empty (S1), Punching (S2), Kicking (S3), Grabbing (S4), Sitting Down (S5), Standing Up (S6), Rotating (S7), Walking (S8), Sitting to Walking (S9), Walking to Sitting (S10), Falling to Walking (S11), and Walking to Falling (S12) \cite{Gao3}. Two datasets are constructed using open-source data (OSSet) and our own recorded data (RWSet), each containing $300$ groups per class. The data is randomly cut in $8:2$ ratio for training and validation. The OSSet is compiled by searching and curating multiple free works within the field \cite{MoVi, NTURGB+D, Kinetics-400, SimHumalator, Charades}. Both OSSet and RWSet are divided into groups of approximately $4$ seconds each. In addition, the hyperparameter settings for network training are shown in TABLE \ref{Training Settings}.\par
\begin{table*}[!ht]
\begin{center}
\caption{PSNR Comparison between Maps Generated by OSSet and From Actual Measurements under Identical Radar Settings$^{*}$.\label{OSSet Comparison}}
\vspace{-0.4cm}
\resizebox{\textwidth}{!}{
\begin{tabular}{cccccccccccccc}
\hline\hline
\multirow{2}{*}{\textbf{Name of the Maps}} & \multicolumn{13}{c}{\textbf{OSSet} ($\mathrm{dB}$)} \\
\cline{2-14}& \textbf{S1} & \textbf{S2} & \textbf{S3} & \textbf{S4} & \textbf{S5} & \textbf{S6} & \textbf{S7} & \textbf{S8} & \textbf{S9} & \textbf{S10} & \textbf{S11} & \textbf{S12} & \textbf{Average among Activities}$^{1}$  \\
\hline
\multicolumn{14}{c}{\textbf{Free-Space Scenario}}\\
\hline
RTM  &$42.15$	&$38.72$	&$45.33$	&$36.91$	&$43.88$	&$35.29$	&$40.17$	&$44.64$	&$39.52$	&$41.06$	&$37.84$	&$45.91$	&$40.95$
\\
Processed RTM  &$18.52$	&$24.11$	&$15.79$	&$22.68$	&$13.94$	&$25.33$	&$19.86$	&$16.47$	&$23.05$	&$14.71$	&$21.22$	&$17.98$	&$19.47$
\\
DTM  &$20.31$	&$25.88$	&$17.65$	&$24.03$	&$15.21$	&$25.92$	&$21.49$	&$18.01$	&$24.76$	&$16.35$	&$23.17$	&$19.55$	&$21.03$
\\       
\hline
\multicolumn{14}{c}{\textbf{TTW Scenario}}\\
\hline
RTM  &$40.26$	&$37.81$	&$44.19$	&$35.88$	&$42.05$	&$36.74$	&$39.92$	&$45.81$	&$38.14$	&$43.57$	&$36.23$	&$44.99$	&$40.47$
\\
Processed RTM  &$16.73$	&$22.48$	&$14.99$	&$20.15$	&$13.24$	&$24.81$	&$17.36$	&$15.91$	&$21.77$	&$14.02$	&$19.64$	&$16.03$	&$17.76$
\\
DTM  &$18.11$	&$23.97$	&$16.25$	&$22.84$	&$14.78$	&$25.13$	&$19.02$	&$17.56$	&$23.31$	&$15.89$	&$21.08$	&$18.42$	&$19.70$
\\   
\hline\hline
\end{tabular}
}
\end{center}
\footnotesize $^{*}$ This experiment compared the PSNR of maps generated by OSSet with that of real-world maps under identical radar settings and activity categories. All results represent the average of $10$ randomly selected datasets.\\
\footnotesize $^{1}$ Take the average of the PSNR results for all $12$ activities.\\
\vspace{-0.4cm}
\end{table*}\par
\begin{table*}[!ht]
\begin{center}
\caption{PSNR Comparison between Maps Generated by RWSet and From Actual Measurements under Identical Radar Settings$^{*}$.\label{RWSet Comparison}}
\vspace{-0.4cm}
\resizebox{\textwidth}{!}{
\begin{tabular}{cccccccccccccc}
\hline\hline
\multirow{2}{*}{\textbf{Name of the Maps}} & \multicolumn{13}{c}{\textbf{RWSet} ($\mathrm{dB}$)} \\
\cline{2-14}& \textbf{S1} & \textbf{S2} & \textbf{S3} & \textbf{S4} & \textbf{S5} & \textbf{S6} & \textbf{S7} & \textbf{S8} & \textbf{S9} & \textbf{S10} & \textbf{S11} & \textbf{S12} & \textbf{Average among Activities}$^{1}$  \\
\hline
\multicolumn{14}{c}{\textbf{Free-Space Scenario}}\\
\hline
RTM  &$43.52$	&$48.19$	&$41.76$	&$46.93$	&$49.21$	&$40.88$	&$45.34$	&$42.67$	&$47.88$	&$44.15$	&$40.23$	&$46.56$	&$44.78$
\\
Processed RTM  &$28.14$	&$32.77$	&$24.59$	&$35.01$	&$37.82$	&$21.93$	&$29.66$	&$31.04$	&$23.48$	&$36.15$	&$26.79$	&$30.58$	&$29.83$
\\
DTM  &$30.25$	&$35.81$	&$27.46$	&$36.99$	&$37.15$	&$22.58$	&$31.73$	&$33.42$	&$25.03$	&$37.89$	&$28.11$	&$32.96$	&$31.62$
\\       
\hline
\multicolumn{14}{c}{\textbf{TTW Scenario}}\\
\hline
RTM  &$41.82$	&$46.25$	&$40.19$	&$44.76$	&$48.91$	&$41.33$	&$47.02$	&$43.18$	&$45.97$	&$48.05$	&$42.64$	&$47.41$	&$44.79$
\\
Processed RTM  &$22.94$	&$28.63$	&$21.47$	&$27.55$	&$31.89$	&$24.16$	&$30.01$	&$26.72$	&$29.38$	&$33.40$	&$25.80$	&$31.12$	&$27.76$
\\
DTM  &$25.33$	&$30.17$	&$23.82$	&$29.41$	&$34.08$	&$26.79$	&$32.55$	&$28.16$	&$31.90$	&$35.74$	&$27.20$	&$33.68$	&$29.90$
\\   
\hline\hline
\end{tabular}
}
\end{center}
\footnotesize $^{*}$ This experiment compared the PSNR of maps generated by RWSet with that of real-world maps under identical radar settings and activity categories. All results represent the average of $10$ randomly selected datasets.\\
\footnotesize $^{1}$ Take the average of the PSNR results for all $12$ activities.\\
\vspace{-0.4cm}
\end{table*}\par

\subsection{Visualizations}
As shown in Fig. \ref{Visualizations_OpenSource} and \ref{Visualizations_RW}, the data generated by the computer vision module and the radar module proposed in the paper are both visualized. Experiments are conducted in Fig. \ref{Visualizations_OpenSource} using open-source HAR video dataset, where the estimated 3D pose, RTM, processed RTM, DTM, and ridge feature extraction results under several typical activities are compared. Fig. \ref{Visualizations_RW} presents the visualization of data recorded by ourselves.\par
From Fig. \ref{Visualizations_OpenSource}, it can be seen that the estimated 3D poses are close to the real videos in terms of structure and spatial position. The generated RTM and DTM can reflect key information about the motion of human joints. Except for the DTM features that are disconnected in the images, effective time-frequency ridges can all be extracted. The motion trends of human joints in free-space and TTW scenarios are consistent with the video trends. Similar conclusions can be drawn from Fig. \ref{Visualizations_RW}, and the quality of the generated data is not affected by common video recording scenarios. The above conclusions together provide qualitative proof of the effectiveness of the generated data.\par

\subsection{Comparative Verification}
The similarity between simulator-generated spectra and experimental data \cite{Gao6, RadHARSimulatorV1} is first compared. Both the same activity categories and radar parameter settings are employed as the generated data. Peak signal-to-noise ratio (PSNR) is used to measure similarity between maps \cite{PSNR}. Assume that $\mathbf{I^\mathrm{set}}$ is the generated map, $\mathbf{I^\mathrm{real}}$ is the corresponding experimental map, and $\mathrm{Max}_{\mathbf{I^\mathrm{set}}}$ is the maximum pixel of $\mathbf{I^\mathrm{set}}$. Then:
\begin{equation}
\mathrm{PSNR} = 10 \cdot \log_{10}\left(\frac{H\cdot W \cdot \mathrm{Max}_{\mathbf{I^\mathrm{set}}}^2}{\sum_{i=0}^{H-1}\sum_{j=0}^{W-1}[\mathbf{I^\mathrm{set}}_{i,j} - \mathbf{I^\mathrm{real}}_{i,j}]^2}\right),
\end{equation}
where $H,W$ are the height and width of the map, respectively. Generally, PSNR values between $15\sim 30\mathrm{~dB}$ are considered similar, values between $30\sim 40\mathrm{~dB}$ are considered good similarity, and values between $40\sim 50\mathrm{~dB}$ are considered nearly identical.\par
\begin{table}[!ht]
\begin{center}
\caption{Ablation Experiment of the Target Detection Model$^{*}$.\label{Target Detection Ablation}}
\vspace{-0.0cm}
\resizebox{0.48\textwidth}{!}{
\begin{tabular}{cccc}
\hline\hline
\textbf{Methods} & \textbf{Maps} & \textbf{OSSet Average}$^{1}$ & \textbf{RWSet Average}$^{1}$  \\
\hline
\multicolumn{4}{c}{\textbf{Free-Space Scenario}}\\
\hline
\multirow{3}{*}{ACF + GNN$^{2}$} & RTM &$39.12$ &$42.63$ 
\\
 & Processed RTM  &$17.92$ &$27.42$
\\
 & DTM  &$19.75$ &$28.85$
\\   
\multirow{3}{*}{RTMDet + MOSSE$^{3}$} & RTM &$39.83$ &$42.83$ 
\\
 & Processed RTM  &$18.42$ &$27.82$
\\
 & DTM  &$20.08$ &$29.51$
\\  
\multirow{3}{*}{Proposed Method} & RTM &$40.95$ &$44.78$ 
\\
 & Processed RTM  &$19.47$ &$29.83$
\\
 & DTM  &$21.03$ &$31.62$
\\       
\hline
\multicolumn{4}{c}{\textbf{TTW Scenario}}\\
\hline
\multirow{3}{*}{ACF + GNN} & RTM &$37.92$ &$41.65$ 
\\
 & Processed RTM  &$15.78$ &$24.87$
\\
 & DTM  &$17.89$ &$26.87$
\\  
\multirow{3}{*}{RTMDet + MOSSE} & RTM &$39.02$ &$42.45$ 
\\
 & Processed RTM  &$16.48$ &$25.57$
\\
 & DTM  &$18.39$ &$27.37$
\\  
\multirow{3}{*}{Proposed Method} & RTM  &$40.47$ &$44.79$
\\
& Processed RTM  &$17.76$ &$27.76$
\\
& DTM &$19.70$ &$29.90$
\\   
\hline\hline
\end{tabular}
}
\end{center}
\footnotesize $^{*}$ Existing human target detection works are referenced from \cite{TargetDetectionModel}. To ensure the rigor of the experiment, the designs of other modules remain unchanged.\\
\footnotesize $^{1}$ Take the average of the PSNR results for all $12$ activities in $\mathrm{dB}$ unit.\\
\footnotesize $^{2}$ ACF is the abbreviation of aggregate channel feature detector.\\
\footnotesize $^{3}$ MOSSE is the abbreviation of minimum output sum of squared error correlation filtering tracking method.\\
\vspace{-0.4cm}
\end{table}\par
\begin{table}[!ht]
\begin{center}
\caption{Ablation Experiment of the Pose Estimation Model$^{*}$.\label{Pose Estimation Ablation}}
\vspace{-0.0cm}
\resizebox{0.48\textwidth}{!}{
\begin{tabular}{cccc}
\hline\hline
\textbf{Methods} & \textbf{Maps} & \textbf{OSSet Average}$^{1}$ & \textbf{RWSet Average}$^{1}$  \\
\hline
\multicolumn{4}{c}{\textbf{Free-Space Scenario}}\\
\hline
\multirow{3}{*}{CPM + Matching$^{2}$} & RTM &$38.52$ &$42.17$ 
\\
 & Processed RTM  &$17.41$ &$27.15$
\\
 & DTM  &$18.93$ &$28.56$
\\   
\multirow{3}{*}{TCN$^{3}$} & RTM &$39.78$ &$43.51$ 
\\
 & Processed RTM  &$18.65$ &$28.77$
\\
 & DTM  &$20.14$ &$30.29$
\\  
\multirow{3}{*}{Proposed Method} & RTM &$40.95$ &$44.78$ 
\\
 & Processed RTM  &$19.47$ &$29.83$
\\
 & DTM  &$21.03$ &$31.62$
\\       
\hline
\multicolumn{4}{c}{\textbf{TTW Scenario}}\\
\hline
\multirow{3}{*}{CPM + Matching} & RTM &$37.68$ &$41.92$ 
\\
 & Processed RTM  &$15.23$ &$24.79$
\\
 & DTM  &$16.94$ &$26.33$
\\  
\multirow{3}{*}{TCN} & RTM &$39.15$ &$43.08$ 
\\
 & Processed RTM  &$16.59$ &$26.13$
\\
 & DTM  &$18.21$ &$27.85$
\\  
\multirow{3}{*}{Proposed Method} & RTM  &$40.47$ &$44.79$
\\
& Processed RTM  &$17.76$ &$27.76$
\\
& DTM &$19.70$ &$29.90$
\\   
\hline\hline
\end{tabular}
}
\end{center}
\footnotesize $^{*}$ Existing human pose estimation works are referenced from \cite{PoseEstimationModel}. To ensure the rigor of the experiment, the designs of other modules remain unchanged.\\
\footnotesize $^{1}$ Take the average of the PSNR results for all $12$ activities in $\mathrm{dB}$ unit.\\
\footnotesize $^{2}$ CPM is the abbreviation of convolutional pose machine.\\
\footnotesize $^{3}$ TCN is the abbreviation of temporal convolutional network model.\\
\vspace{-0.4cm}
\end{table}\par
In TABLE \ref{OSSet Comparison} and \ref{RWSet Comparison}, the similarity between three kinds of spectra generated by the simulator and the measured spectra is compared across $12$ different activities. From TABLE \ref{OSSet Comparison}, since the primary components in the RTM are the direct wave and the low-rank main echo from the wall, the similarity between the simulator and the measured spectrum is relatively high. The processed RTM and DTM primarily reflect the macro and micro-level characteristics of human joint motions, resulting in relatively low similarity. However, based on the average values of the $12$ activities, the generated data remains valid. Similar conclusions can be drawn from TABLE \ref{RWSet Comparison}. Since RWSet aligns with the scenarios and activities observed in our measured radar spectrum, the similarity calculation results are significantly improved. Overall, the results above demonstrate the validity and effectiveness of the generated data.\par

\subsection{Ablation Experiments}
The human target detection and tracking method, the 3D pose estimation and temporal smoothing method, the radar echo modeling method, and the R/DTM feature enhancing method proposed in this paper are compared to existing methods in this paper as ablation experiments. aiming to prove the rationality of the structure design of the whole simulator. All ablations are conducted using similar methodology to TABLE \ref{OSSet Comparison} and \ref{RWSet Comparison}, with PSNR metrics employed for comparison.\par
As shown in TABLE \ref{Target Detection Ablation}, the proposed RTMDet and GNN-based design is compared to the existing human target detection and tracking methods \cite{TargetDetectionModel}. The designs of other modules remain unchanged. The results indicate that modifying the human target detection method to the machine learning-based aggregate channel feature detector (ACF) approach leads to a decline in generated data quality due to reduced accuracy in bounding box positioning. Similarly, switching the tracking method to the minimum output sum of squared error (MOSSE) algorithm within the correlation filter causes tracking interruptions that degrade generated data quality. The conclusions obtained from simulations of free-space and TTW scenarios are consistent, which means that the design of the target detection method is reasonable.\par
\begin{table}[!ht]
\begin{center}
\caption{Ablation Experiment of the Radar Echo Modeling$^{*}$.\label{Echo Modeling Ablation}}
\vspace{-0.0cm}
\resizebox{0.48\textwidth}{!}{
\begin{tabular}{cccc}
\hline\hline
\textbf{Methods} & \textbf{Maps} & \textbf{OSSet Average}$^{1}$ & \textbf{RWSet Average}$^{1}$  \\
\hline
\multicolumn{4}{c}{\textbf{Free-Space Scenario}}\\
\hline
\multirow{3}{*}{w/o Wall Attenuation$^{2}$} & RTM &$40.95$ &$44.78$ 
\\
 & Processed RTM  &$19.47$ &$29.83$
\\
 & DTM  &$21.03$ &$31.62$
\\  
\multirow{3}{*}{w/o Multipath Effect} & RTM &$40.95$ &$44.78$ 
\\
 & Processed RTM  &$19.47$ &$29.83$
\\
 & DTM  &$21.03$ &$31.62$
\\  
\multirow{3}{*}{Proposed Method} & RTM &$40.95$ &$44.78$ 
\\
 & Processed RTM  &$19.47$ &$29.83$
\\
 & DTM  &$21.03$ &$31.62$
\\       
\hline
\multicolumn{4}{c}{\textbf{TTW Scenario}}\\
\hline
\multirow{3}{*}{w/o Wall Attenuation} & RTM &$34.67$ &$38.41$ 
\\
 & Processed RTM  &$13.56$ &$22.03$
\\
 & DTM  &$14.82$ &$23.49$
\\  
\multirow{3}{*}{w/o Multipath Effect} & RTM &$38.47$ &$42.15$ 
\\
 & Processed RTM  &$16.03$ &$25.54$
\\
 & DTM  &$17.75$ &$27.32$
\\  
\multirow{3}{*}{Proposed Method} & RTM  &$40.47$ &$44.79$
\\
& Processed RTM  &$17.76$ &$27.76$
\\
& DTM &$19.70$ &$29.90$
\\   
\hline\hline
\end{tabular}
}
\end{center}
\footnotesize $^{*}$ To ensure the rigor of the experiment, the designs of other modules remain unchanged.\\
\footnotesize $^{1}$ Take the average of the PSNR results for all $12$ activities in $\mathrm{dB}$ unit.\\
\footnotesize $^{2}$ w/o is the abbreviation of word “without”.\\
\vspace{-0.0cm}
\end{table}\par
\begin{table}[!ht]
\begin{center}
\caption{Ablation Experiment of the Feature Enhancing Model$^{*}$.\label{Feature Enhancing Ablation}}
\vspace{-0.0cm}
\resizebox{0.48\textwidth}{!}{
\begin{tabular}{cccc}
\hline\hline
\textbf{Methods} & \textbf{Maps} & \textbf{OSSet Average}$^{1}$ & \textbf{RWSet Average}$^{1}$  \\
\hline
\multicolumn{4}{c}{\textbf{Free-Space Scenario}}\\
\hline
\multirow{3}{*}{Perona-Malik$^{2}$} & RTM &$40.81$ &$44.62$ 
\\
 & Processed RTM  &$17.55$ &$27.91$
\\
 & DTM  &$19.24$ &$29.75$
\\  
\multirow{3}{*}{KSVD$^{3}$} & RTM &$40.73$ &$44.59$ 
\\
 & Processed RTM  &$18.12$ &$28.34$
\\
 & DTM  &$19.86$ &$30.11$
\\  
\multirow{3}{*}{Proposed Method} & RTM &$40.95$ &$44.78$ 
\\
 & Processed RTM  &$19.47$ &$29.83$
\\
 & DTM  &$21.03$ &$31.62$
\\       
\hline
\multicolumn{4}{c}{\textbf{TTW Scenario}}\\
\hline
\multirow{3}{*}{Perona-Malik} & RTM &$40.31$ &$44.61$ 
\\
 & Processed RTM  &$15.82$ &$25.88$
\\
 & DTM  &$17.53$ &$27.64$
\\  
\multirow{3}{*}{KSVD} & RTM &$40.25$ &$44.52$ 
\\
 & Processed RTM  &$16.49$ &$26.21$
\\
 & DTM  &$18.06$ &$28.03$
\\  
\multirow{3}{*}{Proposed Method} & RTM  &$40.47$ &$44.79$
\\
& Processed RTM  &$17.76$ &$27.76$
\\
& DTM &$19.70$ &$29.90$
\\   
\hline\hline
\end{tabular}
}
\end{center}
\footnotesize $^{*}$ Existing image-based feature enhancing and denoising models are referenced from \cite{EnhancingModel}. To ensure the rigor of the experiment, the designs of other modules remain unchanged.\\
\footnotesize $^{1}$ Take the average of the PSNR results for all $12$ activities in $\mathrm{dB}$ unit.\\
\footnotesize $^{2}$ Classic anisotropic diffusion filtering methods.\\
\footnotesize $^{3}$ KSVD is the abbreviation of K-singular value decomposition.\\
\vspace{-0.4cm}
\end{table}\par
As shown in TABLE \ref{Pose Estimation Ablation}, the proposed HRNet, nearest matching, and Kalman filtering-based design is compared to the existing human pose estimation methods \cite{PoseEstimationModel}. The designs of other modules remain unchanged. Replacing the 2D pose estimation method with a convolutional pose machine (CPM) results in a decline in generated data quality. This occurs because the CPM lacks the powerful multi-scale high-resolution feature representation capabilities of an HRNet, leading to positional shifts in the detected joints. TCN also possesses temporal modeling capabilities, but the absolute 3D positions it estimates for human targets are imprecise, which can also lead to a decline in generated data quality. The conclusions obtained from simulations of free-space and TTW scenarios are consistent, which means that the design of the pose estimation method is reasonable.\par
\begin{figure*}[!ht]
    \centering
    \includegraphics[width=\textwidth]{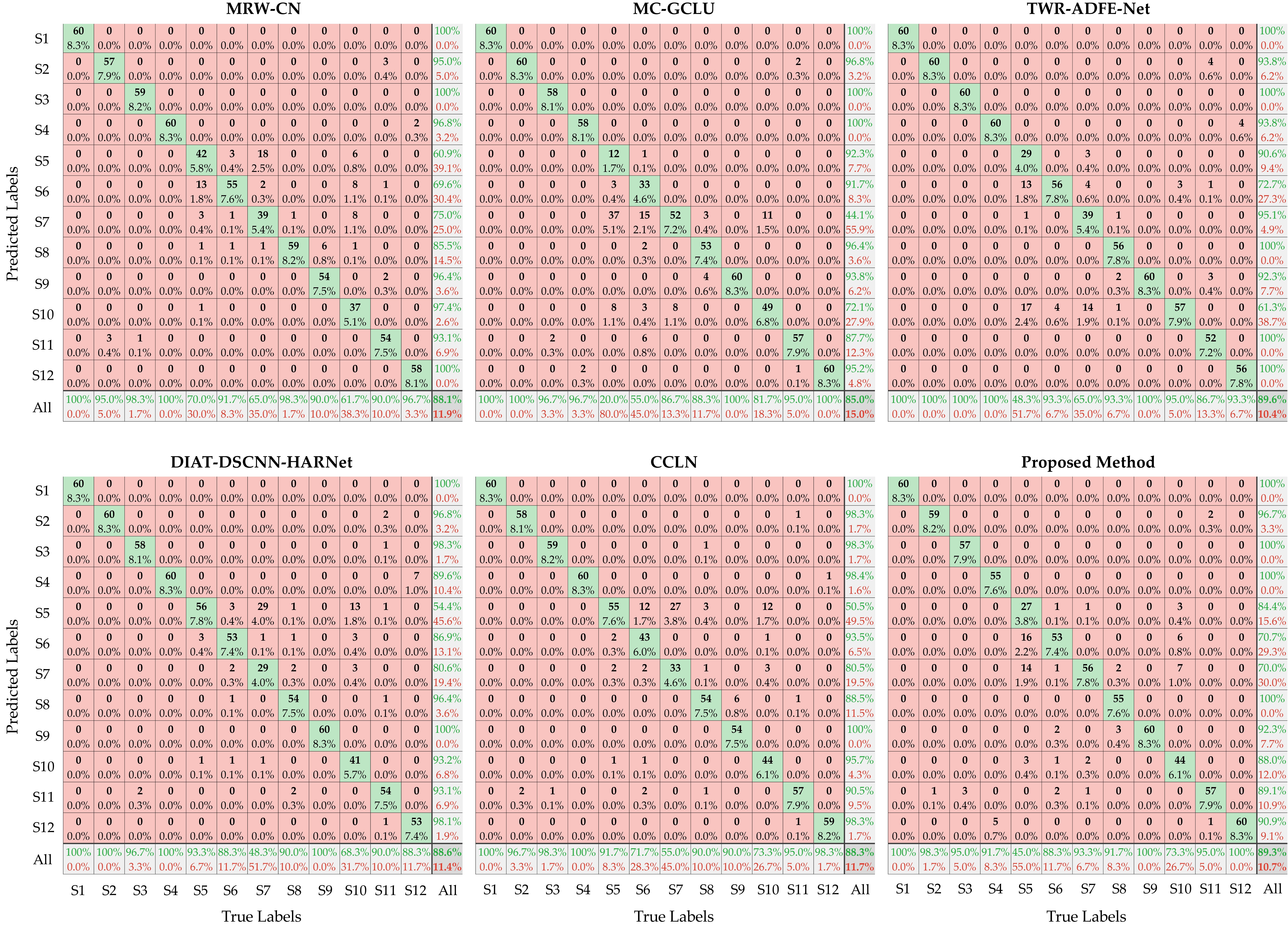}
    \caption{Validation confusion matrices of the proposed method using OSSet compared to existing radar-based HAR methods. “True Labels” represnets the number of actual labels in the validation dataset, and “Predicted Labels” represents the number of labels inferenced by the network model.}
    \label{Confusion_Matrix_OSSet}
    \vspace{-0.1cm}
\end{figure*}\par
\begin{figure*}[!ht]
    \centering
    \includegraphics[width=\textwidth]{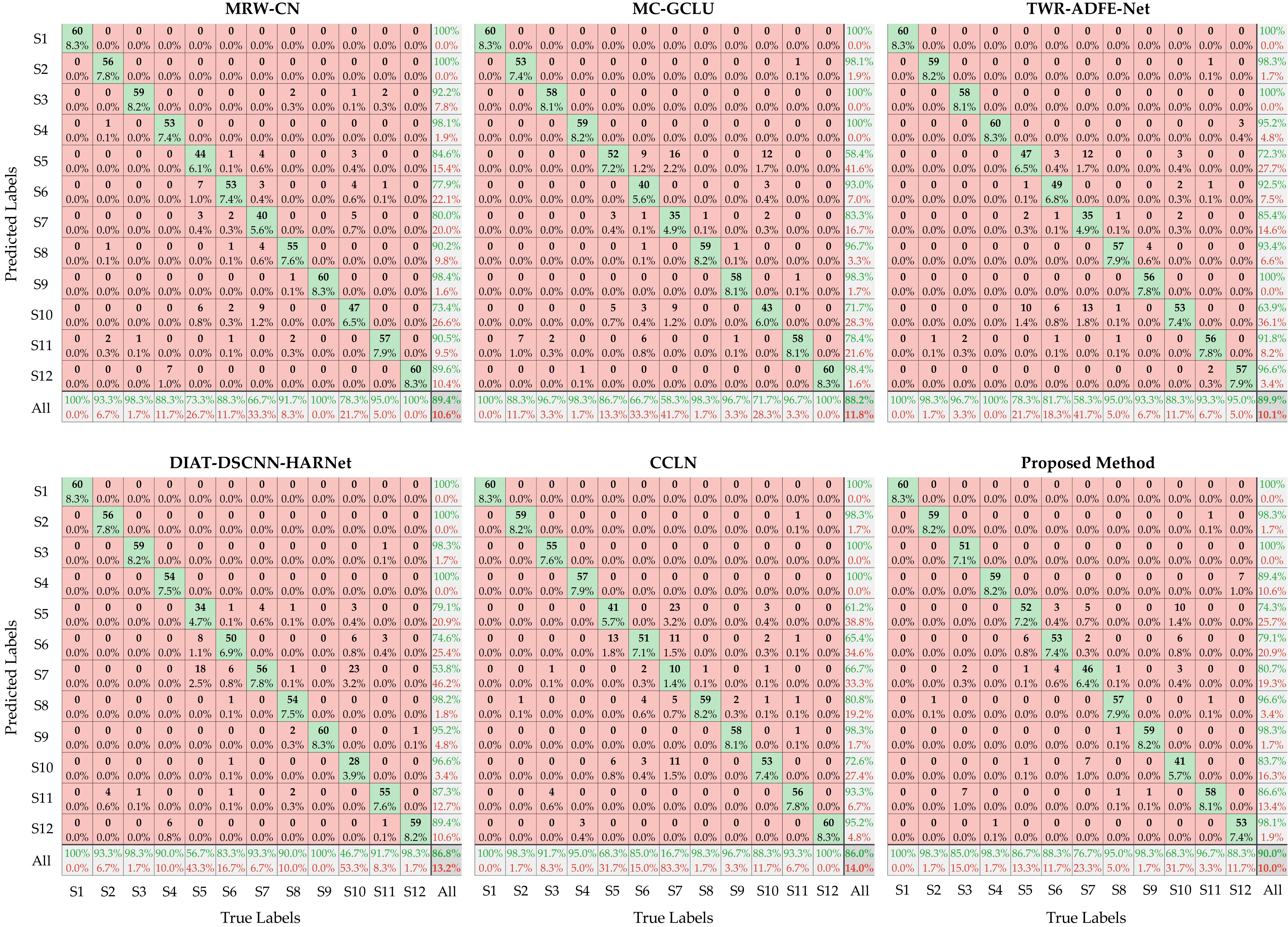}
    \caption{Validation confusion matrices of the proposed method using RWSet compared to existing radar-based HAR methods. “True Labels” represnets the number of actual labels in the validation dataset, and “Predicted Labels” represents the number of labels inferenced by the network model.}
    \label{Confusion_Matrix_RWSet}
    \vspace{-0.1cm}
\end{figure*}\par
As shown in TABLE \ref{Echo Modeling Ablation}, the proposed radar echo modeling design is compared to those without wall attenuation and multipath effect. The designs of other modules remain unchanged. The results of the free-space scenario are thus not affected by this ablation experiment. Based on the experimental results from the TTW scenario, if the mirror-based multipath effect modeling is not enabled, the generated data exhibits differences in micro-Doppler signature structure compared to the measured data, leading to a decline in quality. Without incorporating wall attenuation, the decline in data authenticity becomes more pronounced. The above conclusions collectively demonstrate that the design of the pose estimation method is reasonable.\par
As shown in TABLE \ref{Feature Enhancing Ablation}, the proposed DnCNN-based feature enhancement design is compared to the existing image denoising methods \cite{EnhancingModel}. The designs of other modules remain unchanged. In the results, without employing DnCNN to enhance the radar-generated maps, the quality of features on the RTM shows little variation. This is because the RTM inherently contains almost no effective micro-Doppler information. However, for the processed RTM and DTM, the feature enhancement effect of DnCNN is much more obvious. The conclusions obtained from simulations of free-space and TTW scenarios are consistent, which means that the design of the feature enhancement method is reasonable.\par

\subsection{Network Effectiveness Verification}
To verificate the effectiveness of the radar-based HAR, the proposed method with several existing methods in terms of results such as validation confusion matrices and robustness is compared. Five recognition networks developed based on DTM are selected, including multiscale residual weighted classification network (MRW-CN) \cite{MRW-CN}, masked concatenation with gated convolution and linear unit network (MC-GCLU) \cite{MC-GCLU}, Defence Institute of Advanced Technology radar-based HAR network (DIAT-RadHARNet) \cite{DIAT-RadHARNet}, concatenated CNN-LSTM network (CCLN) \cite{CCLN} for free-space scenario, and TTW radar-based adaptive Doppler feature enhancement network (TWR-ADFE-Net) \cite{TWR-ADFE-Net} for TTW scenario. To ensure the rigor of the experiment, DTM in TTW scenario is used for all network models' training and validation, and all the training hyperparameters are kept consistent.\par
As shown in Fig. \ref{Confusion_Matrix_OSSet} and \ref{Confusion_Matrix_RWSet}, the proposed method is compared with existing methods using confusion matrices derived from network validation. Fig. \ref{Confusion_Matrix_OSSet} shows the validation confusion matrices under OSSet, it can be seen that different network model architectures exhibit varying degrees of sensitivity to different activities' data features. Beyond TWR-ADFE-Net, the proposed method achieves the highest overall validation accuracy among all existing methods while demonstrating strong fitting capabilities for all activities except “Sitting Down (S5)” and “Walking to Sitting (S10)”. Fig. \ref{Confusion_Matrix_RWSet} shows the validation confusion matrices under RWSet, it can be seen that the proposed method achieves the highest validation accuracy among all comparison methods and is the only one with a validation accuracy exceeding $90\%$. Except for the “Walking to Sitting (S10)” activity, the proposed method demonstrated balanced and good fitting capabilities across all other activities. The above results collectively demonstrate the effectiveness of the proposed network model.\par
\begin{figure}
    \centering
    \includegraphics[width=0.48\textwidth]{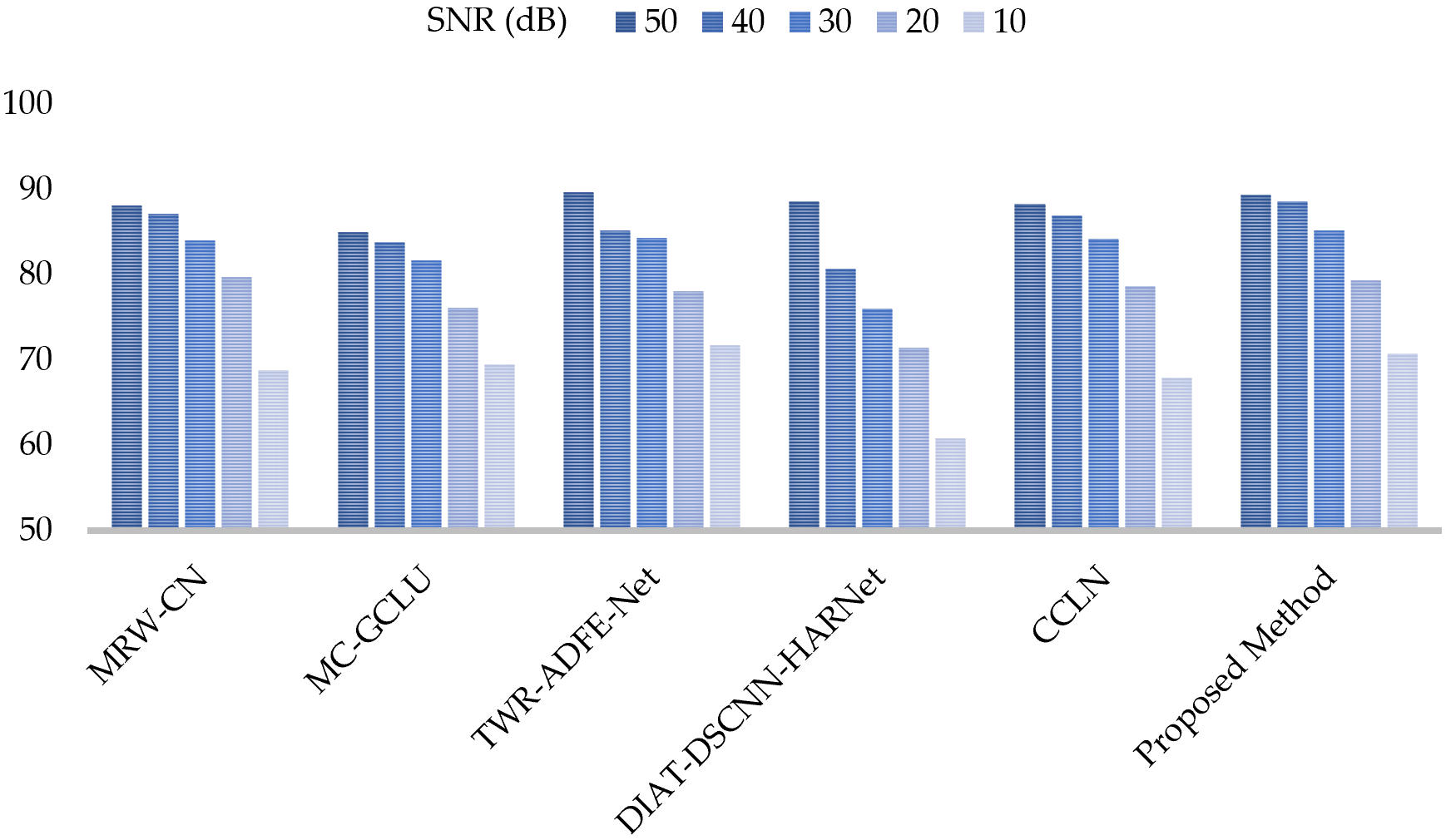}
    \caption{Robustness testing using OSSet.}
    \label{Robustness_OSSet}
    \vspace{-0.0cm}
\end{figure}
\begin{figure}
    \centering
    \includegraphics[width=0.48\textwidth]{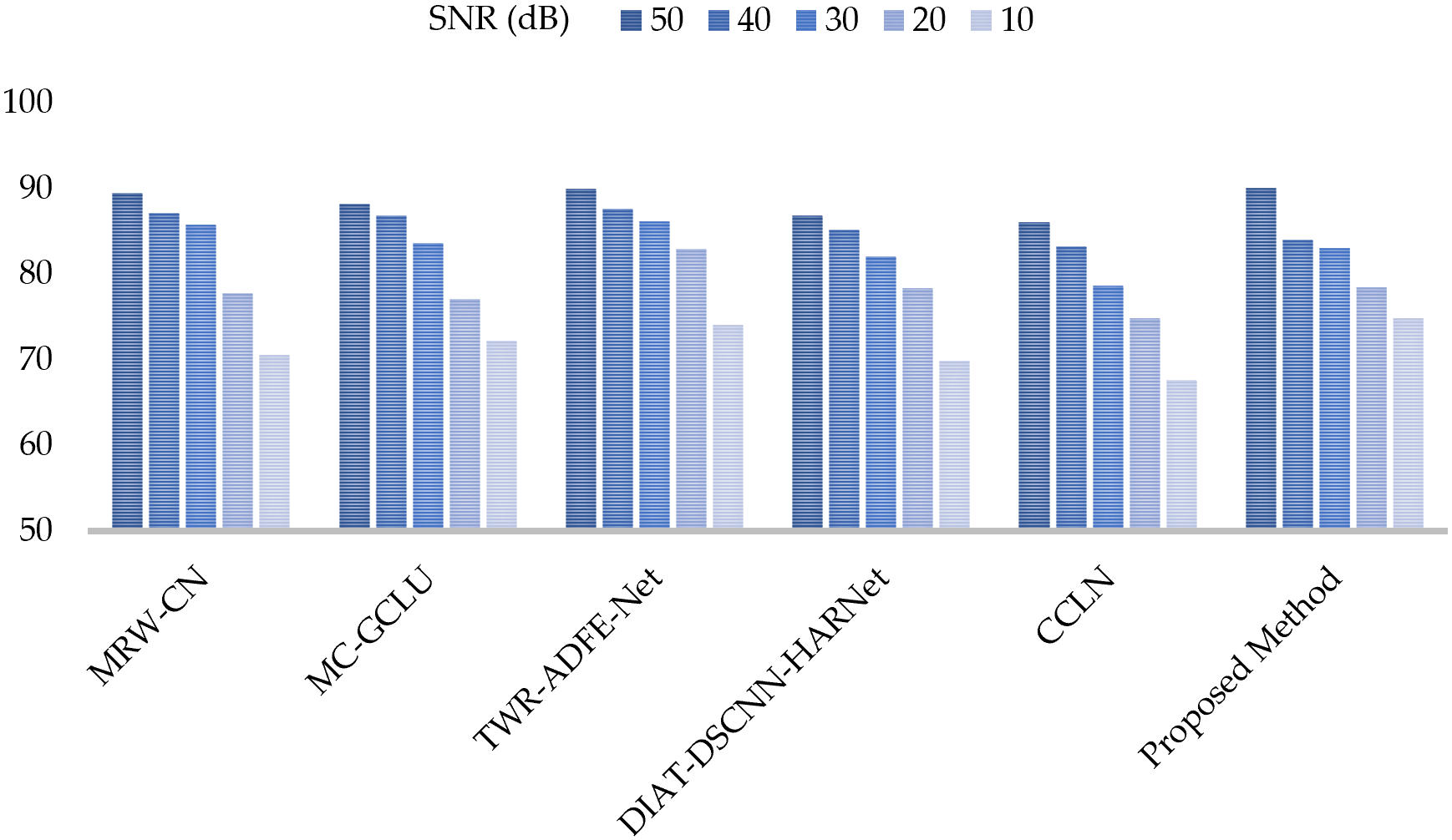}
    \caption{Robustness testing using RWSet.}
    \label{Robustness_RWSet}
    \vspace{-0.2cm}
\end{figure}

As shown in Fig. \ref{Robustness_OSSet} and \ref{Robustness_RWSet}, the proposed method is compared with existing methods using validation accuracy under different simulating SNR settings. Theoretically, as the SNR decreases, the validation accuracy of both the proposed method and existing methods will decline. Therefore, the robustness of the models can be assessed by the magnitude of this accuracy decline that a smaller decline indicates greater robustness \cite{Robustness}. From Fig. \ref{Robustness_OSSet}, when the SNR decreases by $40\mathrm{~dB}$, the proposed method exhibits the least decline in validation accuracy, demonstrating the strongest robustness. From Fig. \ref{Robustness_RWSet}, when the SNR decreases by $10\mathrm{~dB}$, the MC-GCLU demonstrates stronger robustness, while the accuracy of the proposed method declines significantly. As the SNR continues to decrease, the proposed method still exhibits good robustness. The above results collectively demonstrate the effectiveness of the proposed method in resisting noise.\par

\subsection{Discussions}
Although the aforementioned experiments rigorously demonstrated the validity and correctness of the simulator's generated data while validating the proposed radar-based HAR network model, there remains room for improvement in terms of the method's forward-looking design, real-time capability, feature extraction precision, and the physical interpretability of the network architecture. Specifically, this includes the following aspects:\par
\textbf{(1) Forward-Looking and Real-Time Issues:} In the computer vision module of the designed simulator, while the 3D pose estimation method can effectively provide the relative coordinates of human joints, it lacks forward-looking capabilities. Existing monocular camera-based end-to-end 3D pose and location estimation methods can be adopted. Additionally, more lightweight models can be designed to improve inference speed.\par
\textbf{(2) Feature Extraction Issues:} Although ridge features extracted using MLE demonstrate certain effectiveness, experimental results indicate that these features fail when micro-Doppler signature on the simulated maps is discontinuous. More robust feature extraction methods require consideration.\par
\textbf{(3) Network Physical Interpretability Issues:} The designed SPNet method for radar-based HAR is effective. However, when the training data input to the network is changed from image-based DTM to ridge feature, employing a better temporal sequence modeling neural network model or a graph-based neural network model yields superior physical interpretability. This theoretically results in improved accuracy and robustness.\par
\par
\par

\section{Conclusion}
To solve the problem that a comprehensive simulation method for radar-based HAR has still not been developed in this field, and existing software has been developed based on models or motion-captured data, resulting in limited flexibility, RadHARSimulator V2 has been presented in this paper. Both computer vision and radar modules have been included in the simulator. In the computer vision module, the RTMDet with GNN has first been used to detect and track human targets in the video. Then, the HRNet has been used to estimate 2D poses of the detected human targets. Next, the 3D poses of the detected human targets have been obtained by the nearest matching method. Finally, smooth temporal 3D pose estimation has been achieved through Kalman filtering. In the radar module, pose interpolation and smoothing have first been achieved through the Savitzky-Golay method. Second, the delay model and the mirror method have been used to simulate echoes in both free-space and TTW scenarios. Then, the RTM has been generated using pulse compression, MTI, and DnCNN. Next, the DTM has been generated using STFT and DnCNN again. Finally, the ridge features on the DTM have been extracted using the MLE method. In addition, a SPNet architecture has been proposed for radar-based HAR. Numerical experiments have been conducted and analyzed to demonstrate the effectiveness of the designed simulator and the proposed network model.\par

\section{Acknowledgement}
This is the upgraded version of my previous work \href{https://github.com/JoeyBGOfficial/RadHARSimulatorV1-Model-Based-FMCW-Radar-Human-Activity-Recognition-Simulator}{RadHARSimulator V1}. Our simulator is inspired by this interesting work \cite{Text2Doppler}. We extend our sincere thanks for this.\par
I would also like to thank my mentors for the platform they have provided me.\par
My software has not undergone extensive testing by a large number of users. There may still be areas for improvement during use. I welcome your valuable feedback and would be very grateful!\par


\end{document}